# A Revision for the Draconic Gearing of the Antikythera Mechanism, the Eclipse Events of Saros Spiral and their Classification (V.2)

**Aristeidis Voulgaris[1]*, Christophoros Mouratidis[2], Andreas Vossinakis[3]**

[1] City of Thessaloniki, Directorate Culture and Tourism, Thessaloniki, GR-54625, Greece,
[2] Merchant Marine Academy of Syros, GR-84100, Greece,
[3] Thessaloniki Astronomy Club, Thessaloniki, GR-54646, Greece

*Email: arisvoulgaris@gmail.com

## Abstract

*Our research is focused on the missing, but important and necessary Draconic gearing of the Antikythera Mechanism. The three Lunar cycles Sidereal, Synodic and Anomalistic are represented on the Mechanism by correlating the Fragments A and C (part of the Front plate), whereas the fourth Lunar cycle-Draconic results after correlating the unplaced Fragment D with Fragment A. The Draconic scale on the Antikythera Mechanism offers the positional information (ecliptic latitude) of the Moon relative to the Ecliptic. Considering the deformation of the Mechanism's parts during 2000 years underwater and their shrinkage after their retraction from the sea bottom, we present a revised gearing scheme of the Draconic scale. The existence of the Draconic gearing is crucial, because both the preserved and the missing eclipse events can be pre-calculated by the phase correlation of three pointers: of the Lunar Disc, of the Golden sphere/Sun-ray and the Draconic. This means that the eclipse events are calculated by pure mechanical processing and that they are not documented observed events. The phase coordination of the three lunar cycles can be used as a quality criterion for a functional model of the Mechanism. Eudoxus papyrus was the key for the lost words completion of the Back Plate Inscriptions/eclipse events classification of the Antikythera Mechanism.*

## 1. Introduction

The Antikythera Mechanism was a geared device of the Hellenistic era ca. 180 BC. By means of gears, pointers and scales, it displayed the Moon phases, the timed sky path of the Sun and the Moon across the zodiac, it predicted upcoming solar and lunar eclipses with date and hour accuracy and it also showed the starting date of the Athletic Games (Wright 2006; Freeth et al., 2006 and 2008; Anastasiou et al., 2016a; Jones 2017; Seiradakis and Edmunds 2018; Voulgaris et al., 2023b). These calculations are based on the duration (beginning and middle) of the lunar synodic cycle (except the tropical year of the Sun) as it results by the measuring units of the Mechanism's scales (Voulgaris et al., 2025a). Via its pointers and scales, this complex device provided ready-made information on the timing of astronomical and social events, without additional calculations.

The preserved parts of the Antikythera Mechanism incorporate three out of four lunar cycles: Sidereal (Lunar pointer returns to the initial point on the Zodiac Dial ring), Synodic (Lunar pointer aims to Golden sphere-Sun) and Anomalistic (*pin&slot* configuration, Wright 2005; Freeth et al., 2006; Voulgaris et al., 2018b and 2023a). The fourth lunar cycle, the Draconic cycle, named in antiquity as ΑΠΟΚΑΤΑΣΤΑΣΙΣ ΚΑΤΑ ΠΛΑΤΟΣ or ΠΛΑΤΙΚΗ ΑΠΟΚΑΤΑΣΤΑΣΙΣ (restitution in Lunar latitude, Jones 1990; Pirtea 2019; Voulgaris et al., 2022), was also well known and in use during the Hellenistic era:

- Chaldean/(Babylonian) astronomers (Toomer 1984, p.175) had found the equation of 5458 Synodic months = 5923 Draconic cycles, which leads to the recalculation of Saros: 223 Syn. ≈ 241.998 Drac. (instead of integer 242 cycles). This means that after one Saros, the Moon does not exactly returns to the Node[1] (see Voulgaris et al., 2023b in Appendix A, p.62-64)

- Eudoxus Papyrus, written on 165-164 BC (Blass 1887, P.Par. 1, col. XVIII, 10-20) refers to the Lunar Node and its relation to a solar and lunar eclipse.
- Ptolemy in Almagest (Toomer 1984, p.175) mentions that Hipparchus used the equation by Chaldeans (also p.206-209) and in Tetrabiblos he refers to the Moon's position relative to a Node (Ashmand 1900, III, 17, 18).
- In Vettius Valens (Brennan 2022, I, 16)
- Theon of Smyrna (Hiller 1878, p.194)
- Cleomedes (Ziegler 1891, II.5; see also in Spandagos 2002)
- Simplicius (Heiberg 1894, 2.12, p.495).

The Draconic cycle, a very important and critical cycle, which is directly related to the solar and lunar eclipses, that seems to be missing (lost?) from the Mechanism, but can be represented by correlation of fragments A and D (Voulgaris et al. 2022).

There are three specific and critical arguments that lead to the correlation of fragments A and D with an additional gearing with unknown operation (that it could be the Draconic gearing, see **Figure 2**) presented and discussed in Voulgaris et al., 2018b, p.229-230; 2022, p.116-117 and fig. 10; see also 2023b.

- If the Input of the Mechanism was from gear-a1 (taken as *a common sense assumption since 1974*), many mechanical problems regarding the functionality and the handling of the device arise. The Input of the Mechanism from gear-a1 introduces low torque to the following engaged gears (Roumeliotis 2018) and their rotation becomes doubtful and non-seamless. This effect is evident on the *Research Quality* bronze model reconstructions[2].
- Additionally, by starting the Mechanism from a1-gear, the rotation of the Lunar pointer is fast: one tooth (out of 48) rotation of a1-gear, rotates the Lunar pointer by ≈21.3°, i.e. the Lunar pointer runs through ≈70% of a zodiac sign of 30°. In this state, any attempt to aim the Lunar pointer to a desired position is difficult or impossible.

  Here is an equivalent example regarding the steering wheel of a (hypothetical) car: when the driver rotates the steering wheel by (360°/48 teeth) = 7.5°, the wheels of the car rotate by 21.3°. So, driving this car becomes extremely difficult and very dangerous. This car could not pass the test of the driving standards (which usually require that by turning the steering wheel at 7.5°, the wheels rotate 0.5°).
- It is difficult to explain why the ancient Craftsman of the Mechanism would make such doubtful design consideration for its device Input knowing it would create problematic gear motion, risk component damage, present handling difficulties, and offer poor Lunar pointer control.

[1] A good repetition of Saros occurs for the very central eclipses around the Earth's equator and tropics. In every Saros cycle the lunar shadow path gradually moves to northern or southern latitudes and the Saros sequence ends with a partial solar eclipse visible from the North or South Pole, see **Figure A1** in **Appendix-A**. In the same manner, in every Saros cycle, a total lunar eclipse gradually became a partial eclipse and ends as a (small percentage) penumbral eclipse, see **Figure A2** in **Appendix-A**.

[2] The *Protocol of Mandatory Parameters for a Research Quality Bronze Reconstruction of the Antikythera Mechanism*, see Voulgaris et al., 2025a, p.277, Table 6. Reconstructions using modern parts for stabilization, axles/shafts from steel and other modifications change the mechanical status of the device.

In such a way that he could not fully control the Lunar Disc and its pointer, since the precise alignment and positioning of the pointer is very critical to the time calculations it performed (Lunar pointer aims to the Golden sphere-Sun = New Moon, right after a new month begins/a solar eclipse probability, or in opposite position = Full Moon, midmonth/a lunar eclipse probability).

The Antikythera Mechanism is similar to an analog mechanical computer capable for time and events' calculations. The interaction with the User follows a three step scheme:
- The User submits via the Input specific requests to the Mechanism (*Data input*),
- The Mechanism processes the data via its gears (*Processing - the process of transforming input information into and output*) and
- It produces results (predictions) via its pointers and scales (*Output*).
And finally, the User evaluates and uses the results.
In our view, the User and the Mechanism constitute a *Human-Machine System* (Wieringa and Stassen 1999). *Human-Machine System* is a system in which the actions of a human-user and a machine are interrelated, and are both necessary in order to achieve goals and objectives. This interrelation/interaction can continue if the operation and the control of the machine are effective, ergonomic and optimized for handling by the User.
The assumption that the gear-a1 is the Mechanism's Input contradicts the *Human-Machine System* and it is also not compatible to the *Human Body Kinesiology* and *Biomechanics* (Carlton and Newer 1993; Lu and Chang 2010; Duncan et al., 2013; van Bolhuis et al. 1998; Hall 2019).

On the other hand, setting the Mechanism's Input to gear-b3, which is directly connected to the Lunar Disc in ratio 1:1, results in high torque (Roumeliotis 2018; Voulgaris et al., 2018b and 2022), ease of use and a perfect control of the Lunar pointer, which is needed for the accurate pointing. Moreover, as the main measuring unit of the Mechanism is the lunar Synodic cycle (each month of the ancient Greek calendar started right after the New Moon), a close relation between the Input of the Mechanism and the Synodic cycle is necessary. Therefore, for the proper operation of the Mechanism, it is very important to be easy and in precision to aim the Lunar Disc pointer to the Golden sphere-Sun or in the opposite position. This analysis creates a number of impacts, consequences and results that can relate the unplaced Fragment D to a new procedure for the Mechanism (Voulgaris et al., 2022 and 2023b). The existence of the Draconic gearing on the Antikythera Mechanism provides several results, gives answers to many questions, justifies the eclipse events' specific sequence and explains the way of the events' classification. Correlating the generated results by the Draconic pointer and the eclipse magnitudes' description in Eudoxus papyrus (Blass 1887), lead to the reconstruction of the lost Back Plate Inscriptions.

## 2. The deformed and shrunken fragments of the Mechanism

During 2000 years under water, copper of the bronze Mechanism's parts (density 8.8 gr/cm$^3$, alloy of ≈94% Cu and 6%Tin, Price 1974) gradually transformed to a new rocky material named Atacamite [$Cu_2(OH)_3Cl$] (Voulgaris et al., 2019b), which has much lower density (3.8

gr/cm$^3$) and lower absorption in X-Rays (Voulgaris et al., 2018c). Pressure and gravity further deformed the Mechanism's (new material) parts. When the Mechanism retracted from the sea bed the abrupt environment change and exposure to the dry air led to its shrinkage, deformation and cracking (Voulgaris et al., 2019b). Today, most of its parts are broken, shrunk, displaced, deformed, worn out and many flattened parts deviate significantly from flatness, **Figure 1**.

The deformation in the fragments' geometry is stronger to the larger dimension parts and leads to inaccurate dimensional measurements. Dimensional measurement values of a shrunk part are smaller than the original values before the shrinkage: E.g. Budiselic et al., 2020; Woan and Bayley 2024 calculated the total number of holes based on the partially preserved Bearing Base ring on Fragment C (today only an arc is preserved beneath the Egyptian calendar ring), and resulted to the number of 354 holes around the ring, whereas the 365 holes are needed for the functionality of this ring on the Mechanism. The Fragment C presents a strong deformation and shrinkage, evident on its cross section, see fig. 1E in Voulgaris et al., 2025a.

Since deformation and shrinkage renders the dimensional measurement values smaller than the original ones, the measurements by Budiselic et al., 2020; Woan and Bayley 2024 can be used to estimate the percentage of linear shrinkage of the Mechanism's fragments:
*Percentage of shrinkage= part dimension in current condition/original* → 354/365 ≈ 96.7%, and the linear dimension of [shrunk/deformed part] *minus* [original dimension] ≈ –3.3%. The calculated percentage of shrinkage/deformation is not intended as a precise value-rule for the deformation of all fragments. However, we know with certainty that Fragments A and C, are currently deformed and have experienced shrinkage. This value is useful as evidence for the approximate estimation of the original dimension of some partially preserved parts, since today this device does not exist in its original bronze form and dimension.

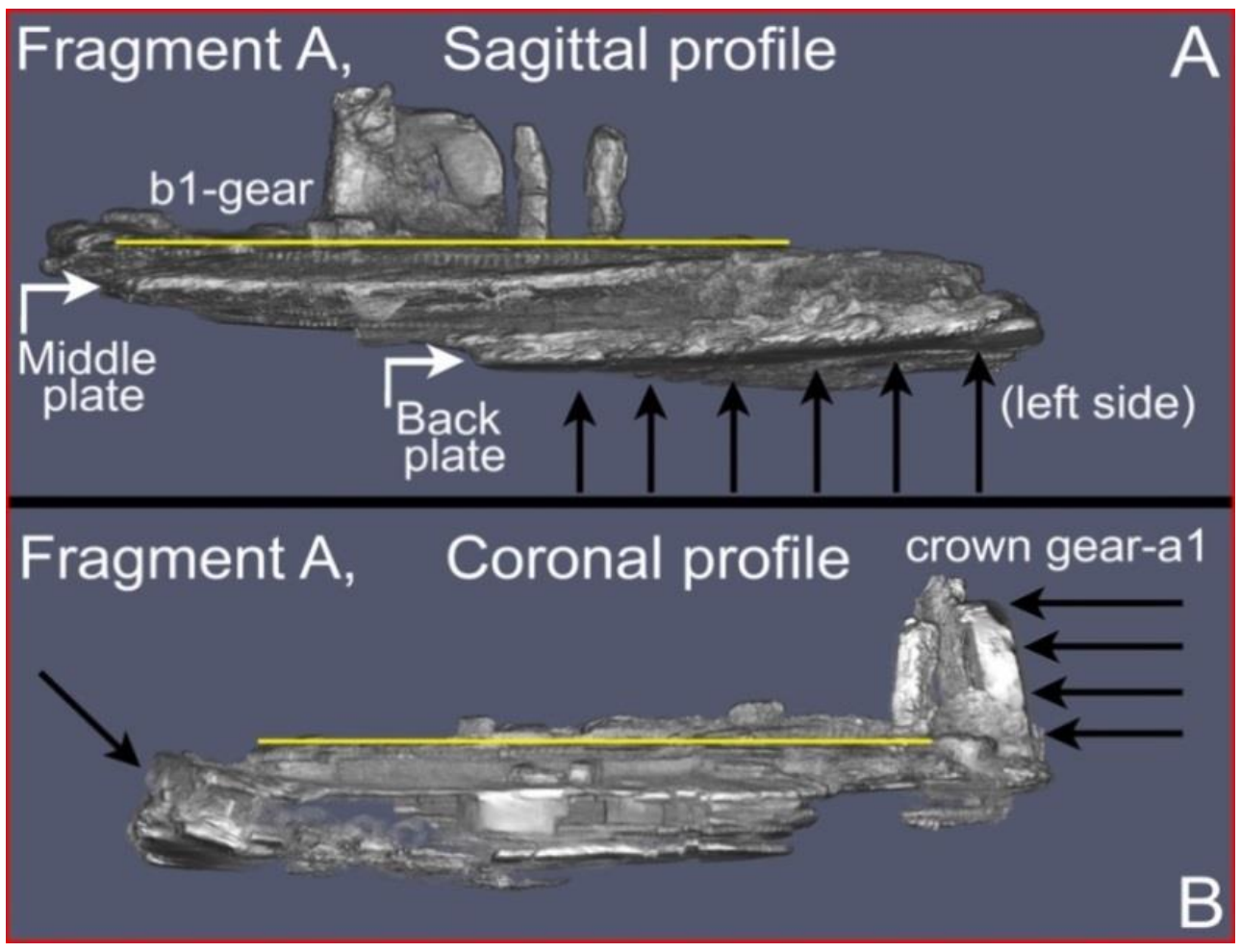

**Figure 1:** A 3D reconstruction of Fragment A from the AMRP RAW Volume data of X-Ray tomographies. The images are oriented to the b1 gear plane (which is just below the yellow horizontal line). The distorted Back and Middle plates should be parallel to the gear-b1. **Panel A:** The sagittal profile of Fragment A. The deformation of Fragment A (black arrows) at its left side is evident.
**Panel B:** The coronal profile of the fragment. The crown gear-a1 deviates from the perpendicularity relatively to gear-b1. These geometrical deviations (bending and torsion) are a result of the shrinkage, displacement and deformation of the parts. Processed images by the authors using *Real 3D VolViCon* software.

The gear-b1 is the largest gear of the Mechanism and represents the daily timed travel of the Sun on the Zodiac sky (Voulgaris et al., 2018a). The gear is partially preserved; the radius of the gear is not constant across its perimeter and many teeth are missing (see graph of Figure 8 in Voulgaris et al., 2022). Gear-b1 isn't a solid disc (as gear-e3 is) but consists of a ring and four arms, and it was probably constructed by scrap bronze parts. Today, gear-b1 isn't flattened, as this non-solid material construction is more prone to shrinkage and distortion. Taking into account the shrinkage of the Mechanism parts, the total number of teeth for the original bronze gear-b1 should be larger than the current measured values. Adopting a value of about 3% of shrinkage, **Table 1** gives the most probable values for the original gear teeth of gear-b1.

**Table 1:** Estimation of the total teeth number of teeth for gear b1 taking into account a value of 97% of deformation/shrinkage. Second column presents the different estimates by several researchers (data taken from Freeth et al., 2006, Supplementary Notes). The third column gives the number of gear-b1 teeth by assuming a correction of ×103% on the shrunk/deformed gear-b1 (left value) and the calculated minimum number of teeth resulting from the present condition of gear (right value).

| Measurements by | Estimated number of gear-b1 teeth measured on the current condition of Fragment A | Probable number of gear-b1 teeth introducing the (estimated) correction of the shrinkage ×103% (left value) |
|---|---|---|
| C. Karakalos | 223–226 | 233(+) – (223) |
| M.T. Wright | 216–231 | 238(+) – (216) |
| D.S. Price 1974 | 225 | 232(+) – (225) |
| Freeth et al., 2006 | 223–224 | 231(+) – (223) |
| Voulgaris et al. 2022 | 219–225 | 232(+) – (219) |

## 3.1 A Revised Draconic gearing for the Antikythera Mechanism

The coordination of the solar tropical years, the lunar sidereal and synodic cycles in integer number of 19 years = 254 sidereal = 235 synodic cycles, creates the Metonic cycle, which it was used during antiquity in order to unite the solar tropical year with the Lunar year of 12/13 synodic months.

The second important time coordination is that 223 synodic equals 242 draconic lunar cycles and also 239 anomalistic. This is the Saros cycle of 18.029787234 years (value resulting from the Antikythera Mechanism gearing), a duration in which the pattern of eclipse sequence and the geometry repeats with a delay of about 8 hours. This time coordination of the three lunar cycles does not correspond to an integer number of solar tropical years. Moreover, 223 synodic cycles does not correspond to an integer number of sidereal cycles, since 223 synodic cycles equals 241.029787234 sidereal cycles. The sidereal cycle is very important for the Mechanism's calculations, as one full turn of the Lunar Disc (which is the ideal and proper Input for the Mechanism operation, see **Introduction**) corresponds into one sidereal cycle. The time difference between the Draconic-Sidereal cycles is very small (≈2.6h) and for this reason there are no characteristic numbers for the gear teeth of the Draconic gearing (applying the usual number of teeth of the Mechanism's gears, presented below).

Taking into account the probable number of the b1-gear teeth, which should be (slightly) higher than the present measured/estimated value, we present a revised gearing scheme for the Draconic gearing of the Antikythera Mechanism in order to improve its precision. We set 229 teeth for gear-b1 (or 228 for the $2^{nd}$ option), since this is the gear for the Tropical solar year of the Mechanism. Gear-b1 rotates the crown gear-a1 (48 teeth). On shaft-a (partially preserved on gear-a1), the gear-r1 (63 teeth, Freeth et al. 2006, Supplementary Notes) of Fragment D is attached and is engaged to the hypothetical gear-s1 (57 teeth). Then gear-s1 is fixed the gear-s2 (56 teeth) which is engaged with gear-t1 (22 teeth). Finally, the Draconic pointer is attached to the shaft-t (see **Figure 2**).

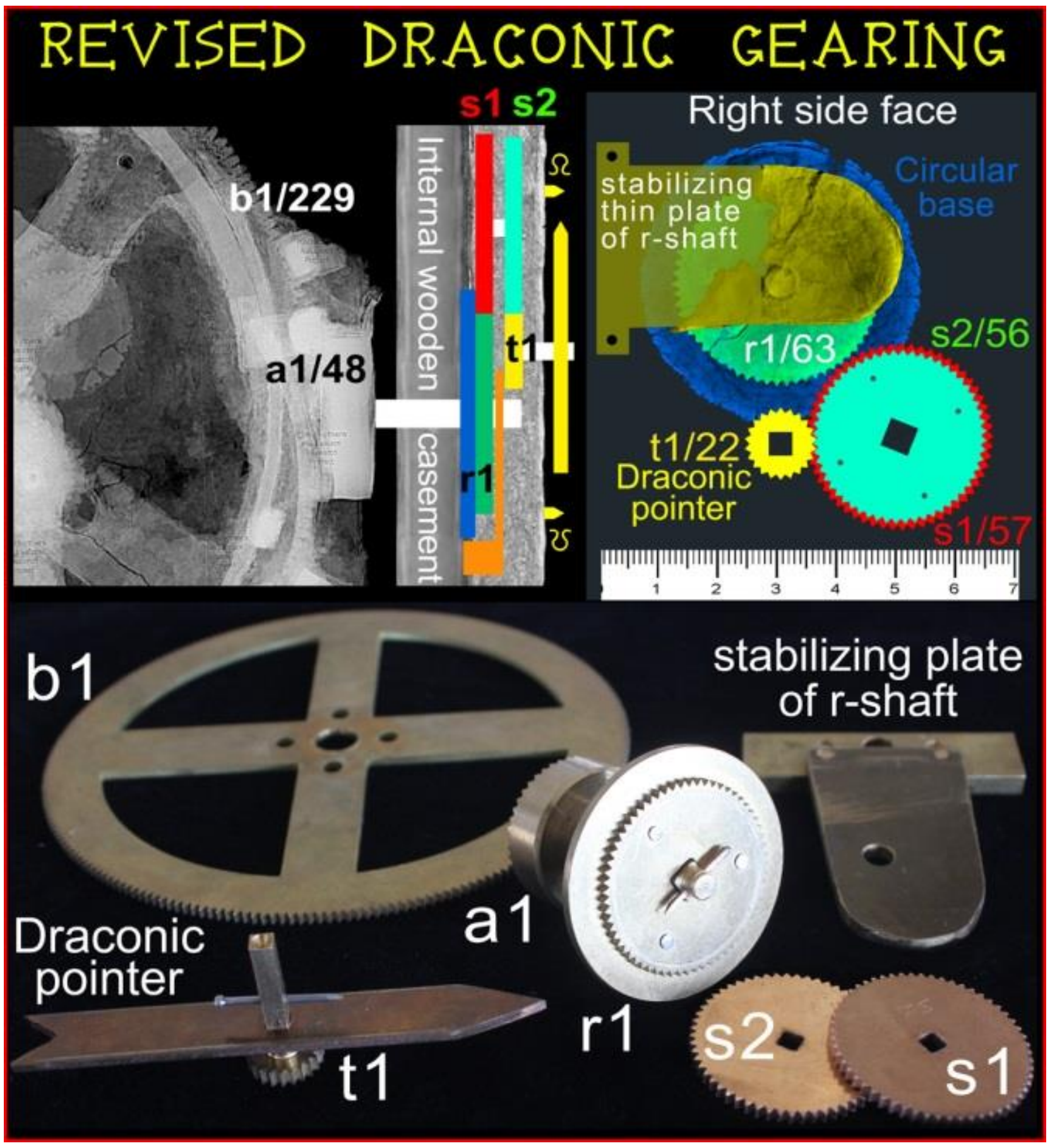

**Figure 2:** Top-left panel, the configuration of the revised Draconic gearing. The annual gear-b1 rotates the crown gear-a1, and afterwards via gears r1, s1, s2, the motion is transmitted to gear-t1. The Draconic pointer is attached to shaft-t. The gearing is located at the right side of the Mechanism inside the External Wooden Casement (Voulgaris et al., 2019b). Top-right panel, the three parts of Fragment D (gear-r1 fixed on its Circular base and the thin stabilizing plate) were processed as separate parts using the corresponding tomographies, and were afterwards aligned and stacked (Voulgaris et al., 2022). The thin stabilizing plate is quite worn out and its remains have collapsed on gear-r1 surface, as the X-ray tomography of the fragment reveals. Most of the Mechanism shafts need to be stabilized between two plates for their proper operation. Bottom panel, the parts of the revised Draconic gearing were constructed in bronze by the authors.

The configuration of the revised Draconic gearing follows:
{223 * (254/235)} * (b3/e1) * (e6/k2) * (k1/e5) * (e2/d2) * (d1/c2) * (c1/b2) * {(b1/a1) * (r1/s1) * (s2/t1)} = 18.029787234 * {(b1/a1) * (r1/s1) * (s2/t1)} =
18.029787234 * {(229/**48**) * (**63**/57) * (56/22)} = 242.0002901 *Equation (1)* or
18.029787234 * {(228/**48**) * (**63**/34) * (61/40)} = 242.0001791 *Equation (2)* turns of Draconic pointer/one Saros.

*Equation (1)* yields an error of +0.0002901 turns of draconic pointer per Saros, corresponds into a pointer's shift ≈0.064476° per Saros (that is equal to one Draconic cycle per 3447 Saros cycles, practically non-detectable error for several decades of Saros cycles, and beyond the time span of the Mechanism operation).

The revised Draconic gearing configuration follows the preserved parts' position (b1, a1), the specific parts of Fragment D and the hypothetical gears s1, s2, t1 according to the constructional characteristics of the ancient Craftsman. This revised gearing scheme meets the dimensional requirements and is fitted to the right side of the Fragment A, between the Internal and the External wooden casement of the Mechanism (Voulgaris et al., 2019b).

The placement of the Draconic scale and pointer at the right side of the Mechanism offers a "*3D projection*" of the Moon relative to the Ecliptic: the Draconic pointer depicts the current position of the Moon relative to the Ecliptic plane-Zodiac Dial ring, (Moon above, on or below the Ecliptic plane) – ecliptic latitude, while at the same time the Lunar Disc pointer on the Zodiac scale depicts the ecliptic longitude, see **Figure 3** and **4**.

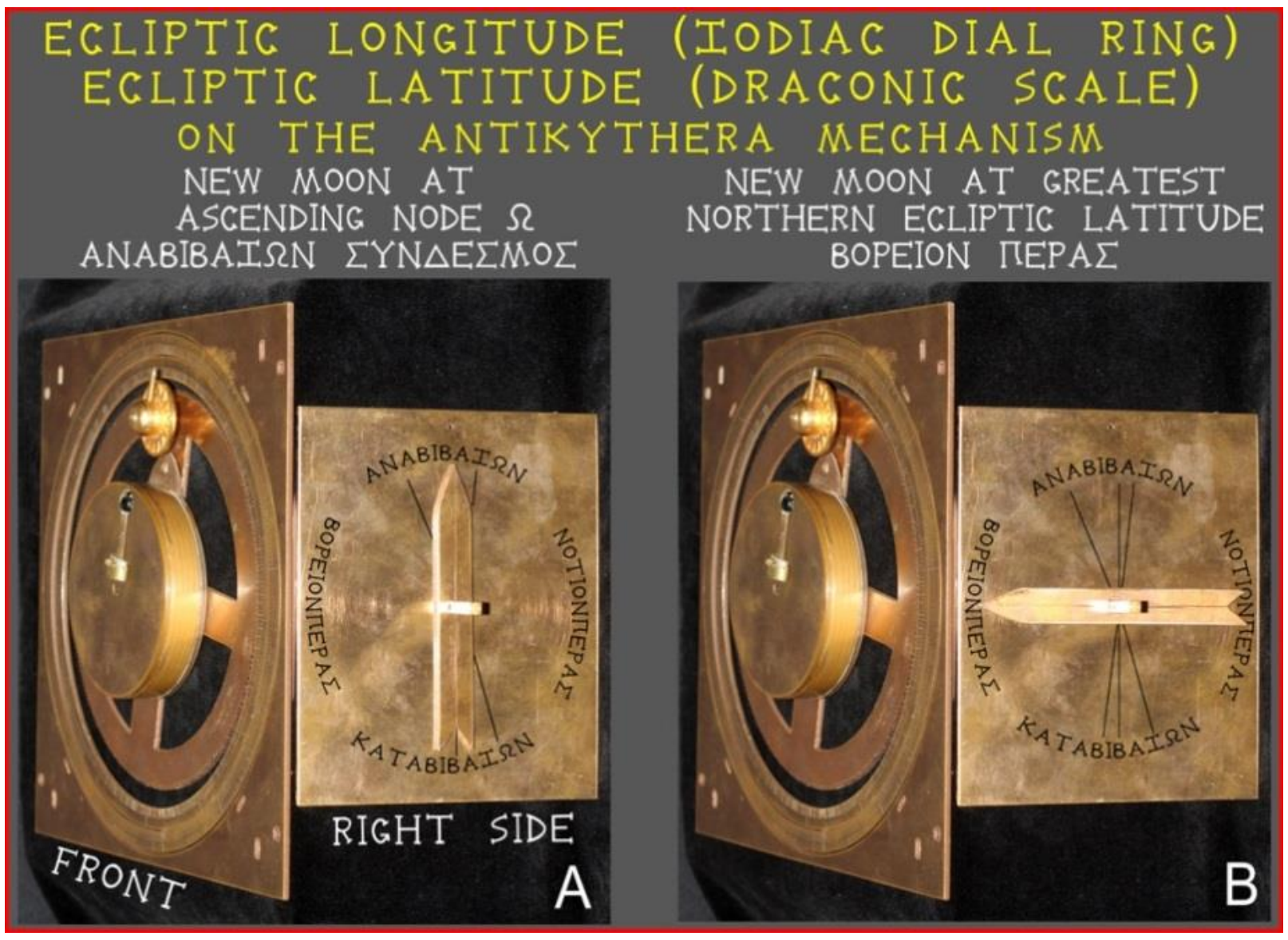

**Figure 3:** The geometrical relation between the Ecliptic plane – Ecliptic longitude (which is defined by the Zodiac Dial ring of the Mechanism) and the Draconic scale/pointer's position - Ecliptic latitude, at the right side of the Mechanism. The Lunar Disc pointer aims to the Golden sphere-Sun (Bitsakis and Jones 2016b). When the Draconic pointer aims to the Ascending (ΑΝΑΒΙΒΑΖΩΝ, **Panel A**) or to Descending (ΚΑΤΑΒΙΒΑΖΩΝ) Node (0, π phase of Draconic cycle), is parallel to the Zodiac Dial calendar ring and then the New Moon is located exactly on the Zodiac ring plane (ecliptic latitude 0). When the Draconic pointer is perpendicular to the Line of Nodes (π/2 or 3π/2 phase of Draconic cycle), i.e. in greatest Northern (ΒΟΡΕΙΟΝ ΠΕΡΑΣ, **Panel B**) or in Southern ecliptic latitude (ΝΟΤΙΟΝ ΠΕΡΑΣ), and then the New Moon is 5.15° above or below the Ecliptic plane/Zodiac Dial ring and is located above or below the Sun. In this way, the "*3D projection*" of the Moon is represented by the Antikythera Mechanism. Bronze parts' designed/constructed and images by the authors.

The existence of the Draconic gearing and scale on the right side of the Mechanism's box offers the following advantages and indirect information for the User:

1) It shows the Moon's real position relative to the Ecliptic plane: on the Ecliptic = at the Node, or out of the Node or at the ecliptic limit or away of the Ecliptic - North/South),

2) It gives the User additional information related to the Zodiac month ring: a) when Node-A is at constellation x, then the Node-B is at the diametrically opposite constellation.[3] This way, the User can find the current position of the lunar orbital plane in the sky. b) When the Full Moon is at its greatest northern ecliptic latitude (maximum Declination) and the Sun is located between Sagittarius – Aries (end of Autumn – begin of Spring), then the Moon reaches its maximum altitude of ≈78°-82° for regions in Greece and it lights up the sky all night. This information could be very useful for military or navigation operations or night transportations or hunting.

[3] The astrologer Vettius Valens (born in 120 AD) in Anthologies 1.16 (Ἀναβιβάζοντα ἀπὸ χειρὸς εὑρεῖν, A Handy Method for Finding the Ascending Node) describes the way to find the two Nodes in the sky versus date (Brennan 2022). He also describes the *Hipparcheion*, a method for calculating lunar positions (1.19).

3) It informs the User of an impending eclipse: When the Lunar Disc pointer aims to the Golden sphere-Sun (or in the opposite position) and the Draconic pointer aims between the ecliptic limits of the Draconic scale, the User can conclude that a solar (or Lunar) eclipse will occur. This is a sign to the User to look at the Back plate of the Mechanism: by observing the cell in which the Saros pointer aims (Anastasiou et al., 2014), he can read additional information about the eclipse event:

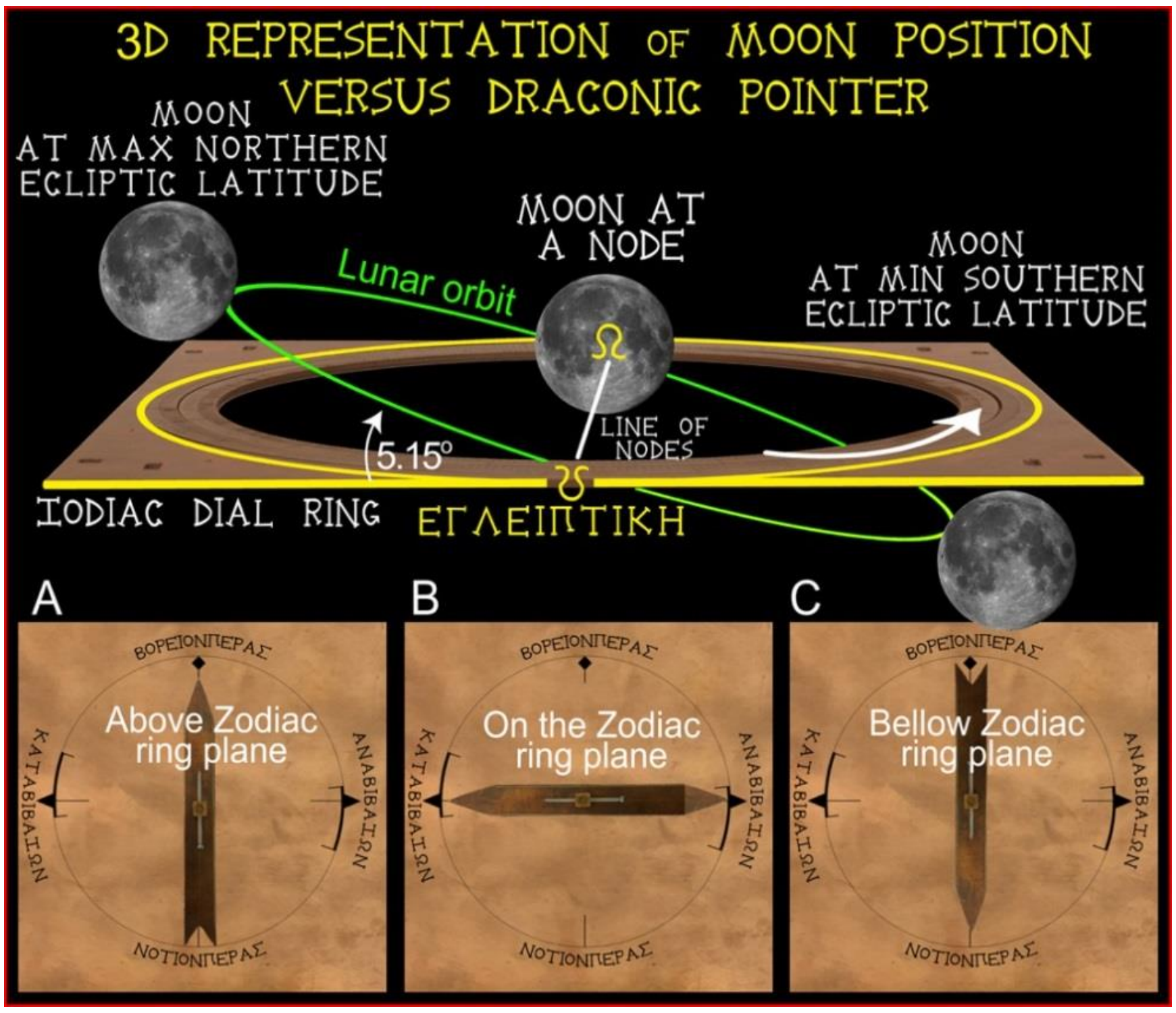

**Figure 4:** The geometrical equivalent in 3D representation of the Moon's position relative to the Zodiac Dial ring/Ecliptic plane, based on the position of the Draconic pointer. When the Draconic pointer aims to the Node (ΑΝΑΒΙΒΑΖΩΝ ☊ or ΚΑΤΑΒΙΒΑΖΩΝ ☋, see **Panel B**), the Moon is just on the Zodiac Dial ring/plane. When the Draconic pointer aims to the ΒΟΡΕΙΟΝ/ΝΟΤΙΟΝ ΠΕΡΑΣ (max Northern/max Southern Ecliptic passage of Moon), the Moon is +5.15°/–5.15° from the Ecliptic plane, see **Panels A** and **C**). During the operation of the Mechanism the (imaginary) Line of Nodes (white line) moves westward, i.e. it rotates CCW in the Zodiac Dial ring (white arrow).

the hour of the event and the direction of the shadow, since the ancient Craftsman has classified the events according to the position of the Moon relative to a Node, which leads to the direction of the Moon projection on the solar disc and the eclipse magnitude of the event (Freeth 2014 and 2019; Anastasiou et al., 2016; Pakzad 2018; Iversen and Jones 2019), see **Section 6**. A relative eclipse events classification is in use today (Pogo 1937).

### 3.2 A Second option for the Draconic gearing and scale

If we make an assumption that the ancient Craftsman did not separate between the ascending and descending nodes, but was only interested to the relative lunar position with respect to the nodes and the greatest ecliptic latitude (max Northern/Southern), then a second option for the Draconic scale can be deduced from *Equation (2)*: By changing the last gear t1 into 20 teeth, the Draconic pointer rotates two times faster than the result from *Equation (2)*:

18.029787234 * {(228/**48**) * (**63**/34) * (61/<u>20</u>)} = 2 * 242.00017912 = 484.0003582 turns of Draconic pointer per Saros, *Equation (3)*.

In this way, the Draconic scale has only one (common) position-point for the two Nodes, one common arc for the ecliptic limits having a double central angle, and one common point for the Maximum/Minimum ecliptic latitude of the Moon (North or South), see **Figure 5**.

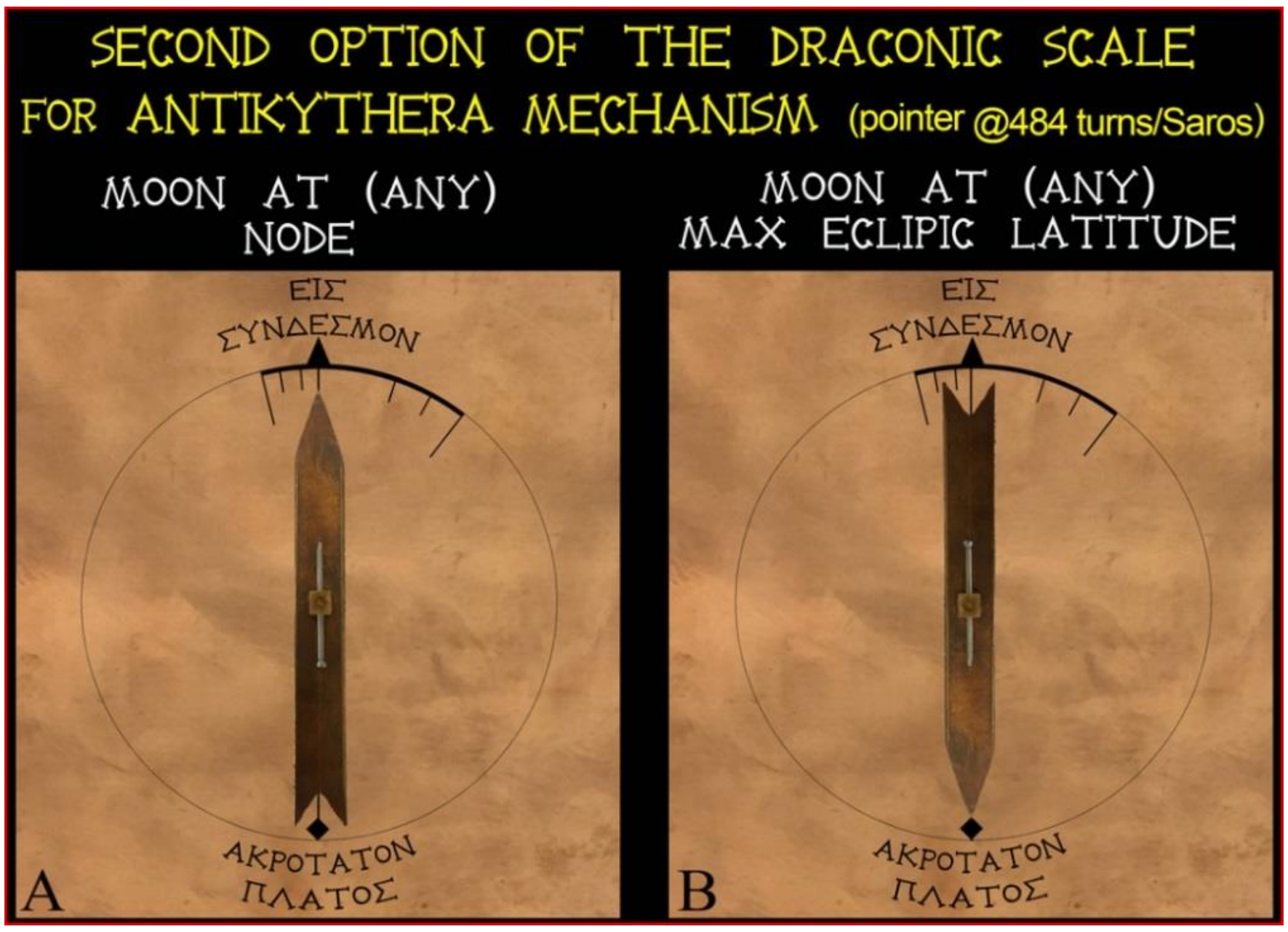

**Figure 5:** The second option for the Antikythera Mechanism Draconic scale. The Draconic pointer rotates two times faster (484 turns per 223 Synodic turns of the Lunar pointer). The position for the two Nodes is common (ΕΙΣ ΣΥΝΔΕΣΜΟΝ – At a Node, **Panel A**), as it is also for the greatest Northern/Southern ecliptic latitude (ΑΚΡΟΤΑΤΟΝ ΕΓΛΕΙΠΤΙΚΟΝ ΠΛΑΤΟΣ, **Panel B**). The Ecliptic limits are represented in one common arc having values ×2 of the initial ecliptic limits angles and they are divided in six unequal sectors for the easier classification of the eclipse events, see **Section 6**. The Nodes-A and B come in sequence (Node-A in an odd number and Node-B in an even number of Draconic pointer turns).

This design offers a better resolution/double magnification of the Draconic pointer’s position between the ecliptic limits and makes the eclipse events classification easier (but it is more sensitive to the gearing positioning errors).

## 4. Detecting the mechanical errors on the Antikythera Mechanism functional models – A mechanics quality criterion for their operation

The Antikythera Mechanism is a unique geared measuring device. As it is a mechanical instrument, its study is also a subject of the *Instrumentation of the geared systems* which analyses the parameters, the behavior and the effects of the mechanical components of a geared device (Voulgaris et al., 2023b).

All geared devices suffer endogenous mechanical errors, **Figure 6**. The mechanical errors on the equatorial mounts of telescopes, on theodolites, on clocks, on the wavelength scales of spectrographs affect the operation, the measurements and the efficiency of these instruments. A general rule is that "*The final results which are calculated by a measuring instrument are affected by the instrument itself*". Geminus in *Introduction to the Phenomena* (18.14) describing the calculation of the mean angular velocity of the Moon writes: Λοιπὰ ἄρα ἐστὶ τὰ ἐκφυγόντα τὴν τῶν φαινομένων διὰ τῶν ὀργάνων παρατήρησιν μιᾶς μοίρας πρῶτα ἐξηκοστὰ κα′ καὶ δεύτερα ι′ (*the rest fractional part of 0° 21′ 10″, apparently escapes observation by instruments*, see Manitius 1898; Spandagos 2002; Bowen and Goldstein 1996). The errors define the final limits on the precision of an instrument. A better quality instrument presents smaller errors, i.e. smaller deviation from the theoretical calculations, resulting in a better approach to Reality. The errors define the final limits on the measurement process of an instrument, because they "*blur*" the results as a lens with aberrations projects the image of an object with low sharpness and contrast (Hecht 2015). The main mechanical errors of the geared devices are the gears' and axes' eccentricities and the gear teeth random non-uniformity (Herrmann 1922; Muffly 1923; Voulgaris et al., 2023b). The eccentricity error happens when the rotation axis (of a gear or a shaft or both) does not coincide with the axis of symmetry, i.e. the rotation axis is not at the center of the gear. Therefore, the gear in point-x has a shorter radius and in the opposite point has a larger radius. The eccentricity affects the position of the pointers and the constant angular velocity, e.g. a clock's minute indicator rotates periodically faster for half period (shorter gear radius) and slower for second half period (larger gear radius).

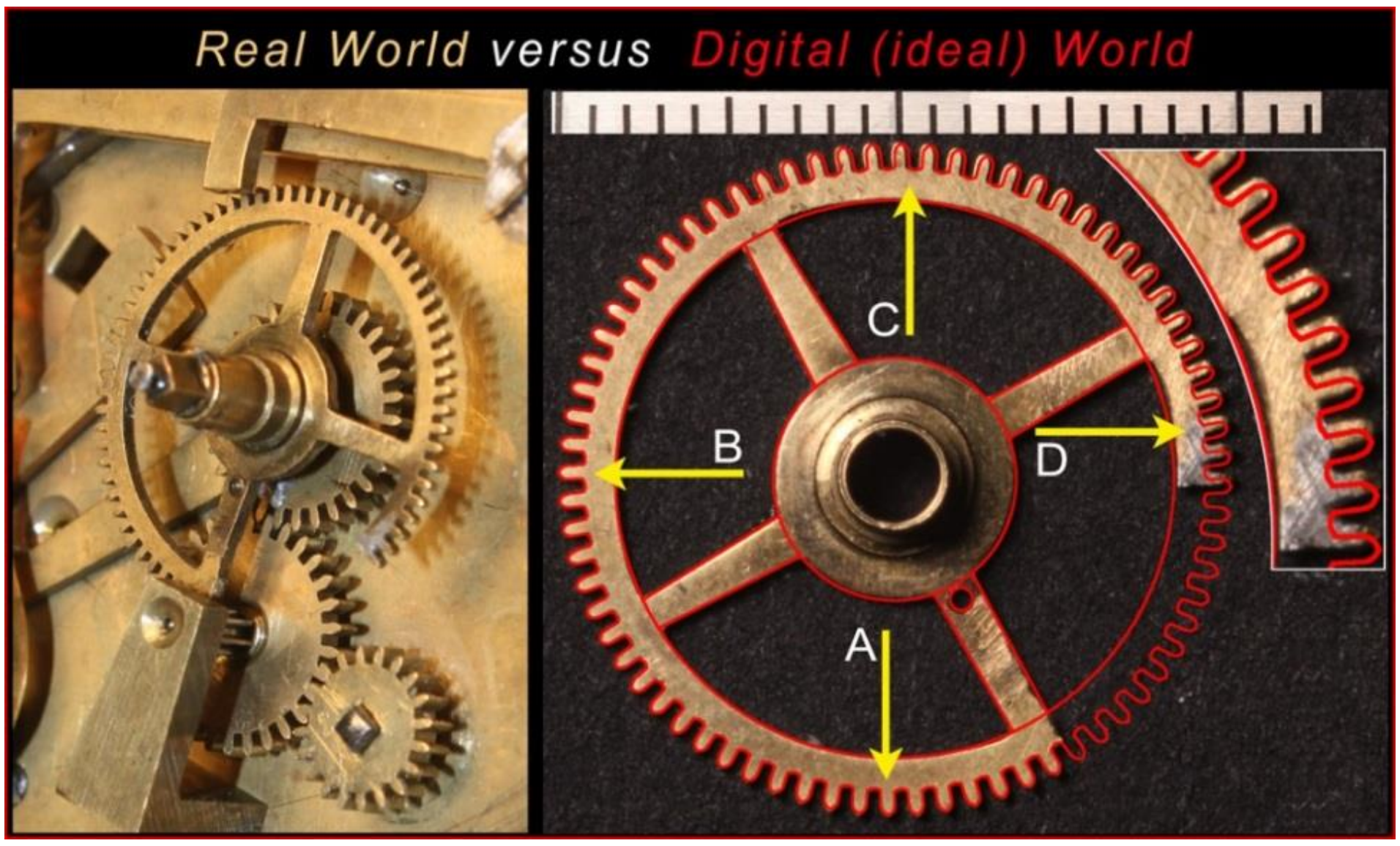

**Figure 6: "*Bronze: the Real World vs Red: the digital ideal world*".** The broken gear of an old clock (it has four arms like gear-b1). The bottom right quadrant broke many times and was repaired using adhesive alloy of tin/lead. The gear has very low eccentricity, but the image reveals that the cause frequent breaking is the result of a division error during the formation process of the 72 teeth, probably because of eccentricity error of the dividing machine of that era (using Computer-Aided Design-CAD software an ideal "digital" gear, having no errors was sketched and aligned exactly on the bronze gear photo (in red color). Arrows A and B indicate positions where there is a perfect match between the real and the digital teeth (zero phase difference). A phase difference gradually appears in arrows C (phase difference ≈π/3) and D (phase difference ≈π/2-half tooth, see insert). Because of the positioning error, the gear teeth were stressed by the steel teeth of the preceding (smaller, higher torque) engaged gear. The image was captured by an optomechanical system designed/constructed by the first author, having low opto-geometrical distortions and in minimized parallax (Voulgaris et al., 2018a).

Endogenous mechanical errors, especially the error of eccentricity or the random non-uniformity on the gear teeth could be present in any of the Mechanism's gears. Eccentricity errors with several values around to 0.1-0.3mm and various directions are present, e.g. on gear m2, with a probable effect detected (see Fig. 8 in Voulgaris et al. 2023b). However, since the preserved parts are deformed and shrunk and other parts are partially preserved or missing, some additional information cannot be recovered. The tooth by tooth motion transmission between engaged gears having triangular shaped teeth also present intermittent ("broken"/non-continuous) motion (see Fig. 7 in Voulgaris et al., 2023b).

The eccentricities and the random non-uniformity of the gear teeth affect the accuracy of the pointers (Edmunds 2011; Voulgaris et al., 2023b and the video *Triangular vs involute gear teeth sound recording* https://www.youtube.com/watch?v=h-qpXYK3bls), causing deviations from the proper/ideal position of the pointers with errors around ±(1°-3°) or ±4°. Much higher positional errors occur on gears that are close to the end of a gearing (i.e. close to the pointers), as result of non-uniform teeth.

Digital 3D simulation can calculate the theoretical perfect/ideal position of the Mechanism's pointers *versus* time since in this case there are no mechanical errors, friction or other constructional mismatches, and the digital gears/pointers operate in an ideal and perfect mechanical world.

These construction errors of the Antikythera Mechanism affect the accuracy of the predicted eclipse events, especially the events located too close - out or inside - the ecliptic limit boundary are affected by the construction errors of the Mechanism (Voulgaris et al., 2023b). Due to the (hypothetical, but very probable) gears' errors, the Draconic pointer deviates from the theoretical position. In marginal case when the theoretical position of the Draconic pointer is inside the ecliptic zone (= eclipse event) and very close to an ecliptic limit, due to the errors of the gearing, the pointer may indicate erroneously out of the limits (= no event) and vice versa (see Figure 10 in Voulgaris et al., 2023b).

This inaccuracy also explains the absence of some eclipse events that should theoretically be calculated (=missing events) and the existence of some other events that should not exist according to the theoretical calculations (=additional events).

The gear(s') eccentricity can also affect the Draconic pointer position relative to the Node: E.g. The pointer should be theoretically located just on Node-X, whereas its real position is before/after (i.e. north or south) the respective Node. Therefore, the classification of the eclipse events (relative to the Node), can be affected by the endogenous gearing errors of the instrument (see **Figures A4-A7** in **Appendix-A**).

Finally, each instrument has its "*personal errors*", (see Figure 4 in Voulgaris et al., 203b). The eclipse events' sequence and classification can (slightly) vary in different instruments, and the results can deviate from the theoretical classification deduced by an imaginary ideal/perfect device, see **Table 6**.

The errors on the pointers' position (Lunar Disc, Golden sphere and Draconic) of a functional reconstructed model of Antikythera Mechanism can be detected and measured by checking the deviation between the (theoretical) position of the pointers during characteristic phase/antiphase time coordination and the pointers' current position in a certain time span. We call this arithmetical resonance as "*The Key Time Points for Mechanism's pointers*" dedicated for Metonic and Saros cycles. This is a mechanics criterion for the quality evaluation of a functional model of the Mechanism, see **Table 2**.

**Table 2:** The arithmetical resonance of the lunar and solar-tropical cycles' in 0, π/2 or π phase, for Metonic and Saros periods, can be used as a quality criterion of a functional model of the Antikythera Mechanism: by checking the pointers' deviation from the theoretical position derived from the arithmetical resonance cycle phase (theoretical position). The following calculations are based on the starting date of the Antikythera Mechanism pointers on 22/23 December 178 BC: New Moon at Apogee (*pin&slot* starts at Apogee), Golden sphere-Sun and Lunar Disc pointer at $1^{st}$ day of Capricorn, Draconic pointer at Node-A (Descending) (Voulgaris et al., 2023a, 2023b and 2023c). In specific cycles, the phase correlation in 0 or π or π/2 (should) appear on the Mechanism pointers. The mechanical errors create small or larger deviations by the cycles' phase coordination.

| **QUALITY CRITERION I:** ***KEY TIME POINTS FOR METONIC CYCLE*** | | |
|---|---|---|
| **Percentage of Metonic cycle** | **Synodic cycle (Lunar Disc pointer) *vs* Solar Tropical year (Golden sphere) Key Time points** | **Sidereal cycle *vs* Solar Tropical year Key Time points (New Moon, Sun start from $1^{st}$ Capricorn)** |
| **Full Metonic cycle $19^{y}$, 235 Synodic, 254 Sidereal** | After 235 repositions of the Lunar Disc pointer to Golden sphere (starting from $1^{st}$ Capricorn), the Sun pointer must return to its starting position, $1^{st}$ Capricorn | After 254 full rotations of the Lunar Disc pointer (starting from $1^{st}$ Capricorn), the Sun pointer must return to its starting position, $1^{st}$ Capricorn |
| **Half Metonic cycle $9.5^{y}$, 117.5 Synodic, 127 Sidereal** | After 9.5 years, the Sun pointer must rotate 9+0.5 turns (= in opposite position to $1^{st}$ Capricorn). 117+0.5 repositions of the Lunar Disc pointer (= opposite position to Golden sphere) | After 127 full rotations of the Lunar Disc pointer (sidereal cycle stars from $1^{st}$ Capricorn), the Sun pointer must rotate 9+0.5 turns (= opposite position to $1^{st}$ Capricorn). Full Moon at $1^{st}$ Capricorn |
| **¼ Metonic cycle $4.75^{y}$, 58.75 Synodic, 63.5 Sidereal** | After 58+0.75 repositions of the Lunar Disc pointer (= Last quarter), the Sun pointer must aim at 90° before $1^{st}$ Capricorn | After 63+0.5 rotations of the Lunar Disc pointer (= opposite position to $1^{st}$ Capricorn), the Sun pointer must aims at 90° before $1^{st}$ Capricorn |
| **¾ Metonic cycle $14.25^{y}$, 176.25 Synodic, 190.5 Sidereal** | After 176+0.25 repositions of the Lunar Disc pointer (= First quarter), the Sun pointer must aim 90° after $1^{st}$ Capricorn | After 190+0.5 rotations of the Lunar Disc pointer (= opposite position to $1^{st}$ Capricorn), the Sun pointer must aim 90° after $1^{st}$ Capricorn |
| **QUALITY CRITERION II:** ***KEY TIME POINTS FOR SAROS CYCLE*** | | |
| **Percentage of Saros cycle** | **Repositions of Lunar Disc pointer to Golden sphere, Synodic cycle *vs* Draconic cycle - Key Time points** | **Repositions of Lunar Disc pointer to Golden sphere, Synodic cycle vs Anomalistic cycle - Key Time points** |
| **Full Saros $18.02978^{y}$ 223 Synodic, 242 Draconic, 239 Anomalistic** | After 223 repositions of the Lunar Disc pointer to Golden sphere, the Draconic pointer (after 242 full rotations) must return to Node-A | After 223 repositions of the Lunar Disc pointer to Golden sphere, the *Pin* (after 239 full rotations of gear-k2) must return at *Apogee* (pin in its largest distance from gear-k2 center |
| **Half Saros - Sar 111.5 Synodic, 121 Draconic, 119.5 Anomalistic** | After 111+0.5 repositions of the Lunar Disc pointer to Golden sphere (= opposite position to Golden sphere), the Draconic pointer must return to Node-A (after 121 full rotations). Full Moon at Node-A | After 111+0.5 repositions of the Lunar Disc pointer to Golden sphere (= opposite position to Golden sphere), the *Pin* (after 119+0.5 rotations of gear k1) must be in the shortest distance from center of gear-k2 (*Perigee*) |
| **¼ Saros 55.75 Synodic, 60.5 Draconic 59.75 Anomalistic** | After 55+0.75 repositions of the Lunar Disc pointer to Golden sphere (= last quarter), the Draconic pointer must aim at Node-B (after 60+0.5 rotations) | **(–)** (the phase of 0.75 Anomalistic cycle it cannot be defined on the *Pin&Slot* system, see Fig. 16 in Voulgaris et al., 2018b) |
| **¾ Saros 167.25 Synodic, 181.5 Draconic 179.25 Anomalistic** | After 167+0.25 repositions of the Lunar Disc pointer to Golden sphere (= first quarter), the Draconic pointer must aim at Node-B (after 181+0.5 rotations) | **(–)** (the phase of 0.25 Anomalistic cycle it cannot be defined on the *Pin&Slot* system see Fig. 16 in Voulgaris et al., 2018b) |

The ancient Craftsman could probably use them to (periodically) check and correct the pointers' position by a minor rotation in a small angle of their corresponding scales (Zodiac month ring and hypothetical Draconic scale). He could also slightly change the direction of the Golden sphere/Sun and its pointer to a specific subdivision (day) of the Zodiac month ring (by slightly aiming the Sun pointer to the correct subdivision).

## 5. Revised ecliptic limits and eclipse events

For the recalculation of the eclipse events we recalibrated *DracoNod(-V2)* visualization program (Voulgaris et al., 2023b) by revising the ecliptic limits[4] in order to improve the best match to the preserved eclipse events, see **Figure 7**.

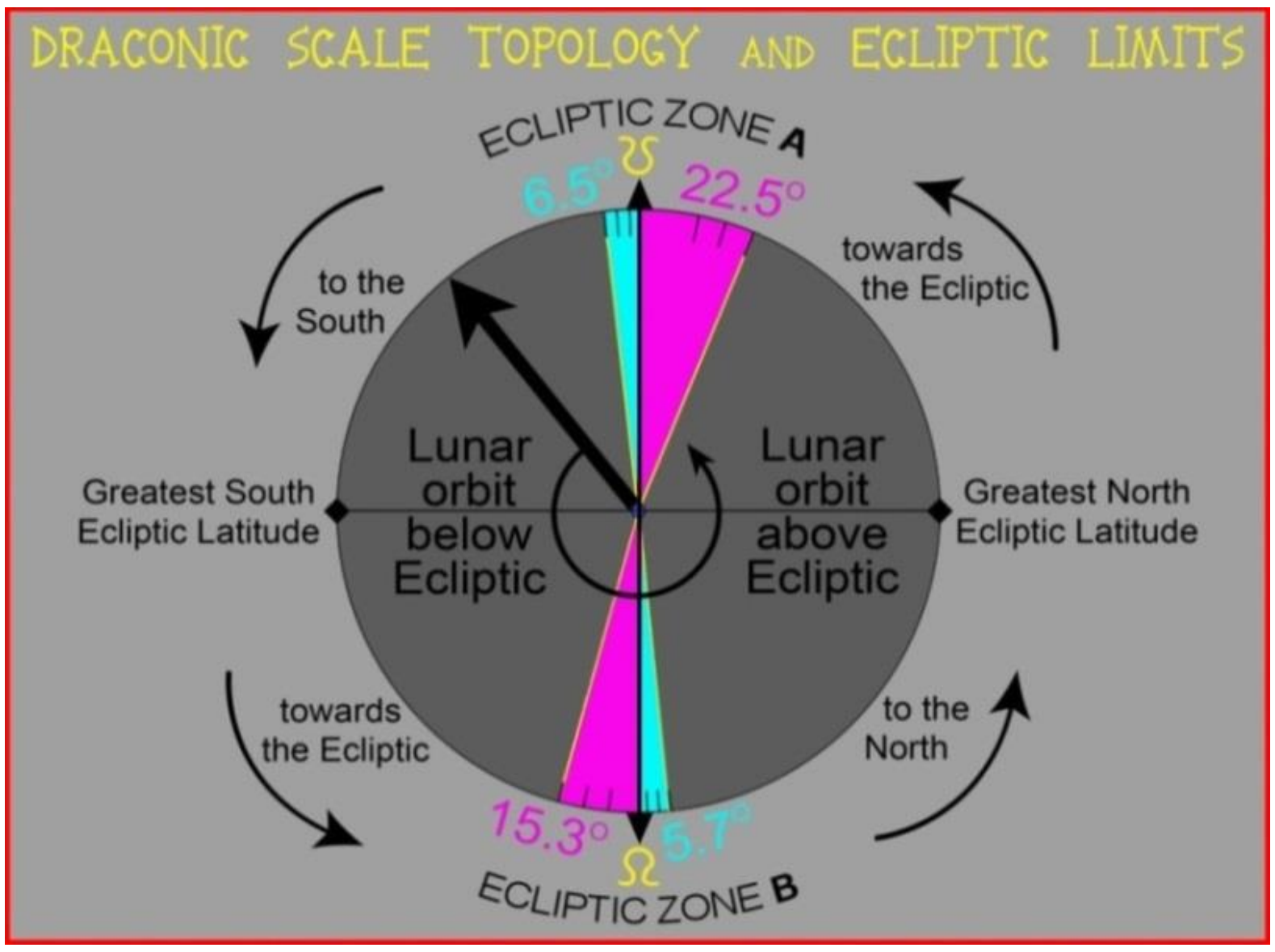

**Figure 7:** The Draconic scale topology. The Line of Nodes divides the scale into the left hemisphere, representing the lunar orbit below/South of Ecliptic and into the right hemisphere - lunar orbit above/North of Ecliptic. The revised ecliptic limits, according to the preserved eclipse events, are also sketched. Each ecliptic zone is divided in six unequal Sectors plus a linear sector(s)-area of Node, related to the eclipse events' classification (see **Section 6**, **Table 5**, **Figures 9** and **10**, and **Figures A4-A7** in **Appendix-A**). (Here the Sectors are presented in a "*deformed*" position due to the gearings' errors).

This revision sheds light to the controversial index cursive letter **ω** (Freeth 2019; Iversen and Jones 2019), which is engraved on the Back Plate Inscription (BPI of Fragment F, L.31), see at the end of **Table 3**.

**Ecliptic Zone A: 6.5° South (CCW) and 22.5° North (CW) from Node-A,**
**Ecliptic Zone B: 5.7° North (CCW) and 15.3° South (CW) from Node-B,** (both include the gearing errors). Mean values: **18.9° before Nodes** and **6.1° after Nodes**), and they are common for the lunar and solar eclipses.

The eclipse events were recalculated via *DracoNod-V2* visualization program (Voulgaris et al., 2023b) by applying the revised ecliptic limits and by using the equation 223 synodic months = 242 draconic months (= 239 Anomalistic months). The revised eclipse events are presented on **Table 3** and in **Figure 11**.

---

[4] The ecliptic limits define the boundary of the Ecliptic zone. When the New Moon/Full Moon is located inside the ecliptic zone, a solar or lunar eclipse occurs, see Green 1985; Smart 1949.

**Table 3:** The revised Saros spiral eclipse events recalculated using the authors' *DracoNod-V2* program (Voulgaris et al., 2023b). A total 64 eclipse events are presented according to the new cell numbering (Voulgaris et al. 2021). The *Ecliptic Zone A/B* (ez-A/B) is defined by the Node: A-☋/B-☊). Preserved letters in bold (Freeth 2014 and 2019; Anastasiou et al., 2016a; Pakzad 2018; Iversen and Jones 2019). *DracoNod-V2* starts with the New Moon at Node-A and at Apogee (Voulgaris et al., 2023a, 2023b, 2023c). The Index letters of some lost solar eclipse events are preserved on the events' classification on the Back Plate Inscriptions (BPI) (Freeth 2014 and 2019; Anastasiou et al., 2016a; Pakzad 2018; Iversen and Jones 2019), and this is a proof of their existence on the cells of Saros spiral (see also **Figure 11**). Regarding the controversial index letter cursive **ω** see at the end of the table.

| Event # | Event index letter (numbered cells) | New cell numbering (Voulgaris et al., 2021) | Preserved Eclipse events on Saros cells | Revised Eclipse events generated by *DracoNod-V2* | Moon position relative to the Node or ecliptic limit/ comments |
|---|---|---|---|---|---|
| 1 <u>Saros begins</u> | [A1] (1) | Cell-1 ΜΕΓΙΣΤΗ ΗΛΙΟΥ ΕΓΛΕΙΨΙΣ | Non preserved Cell | Sun, *Longest Annular Solar eclipse* (ez-A) | **Saros cycle begins. New Moon exactly *at Node-A* ☋ *and* at *Apogee*** |
| 2, 3 | **B1 (2)** (BPI) | **Cell-7** | **Moon, Sun** | **Moon** (ez-A) **Sun** (ez-B) | |
| 4 | **Γ1 (3)** | **Cell-12** | **Sun** | **Sun** (ez-A) | New Moon at ecliptic limit see **Figure A8** |
| 5 | [Δ1] (4) | Cell-13 | Non preserved Cell | Moon (ez-B) | |
| 6 | **E1 (5)** | **Cell-19** | **Moon** | **Moon** (ez-A) | |
| 7 | **Z1 (6)** (BPI) | **Cell-24** | **Sun** | **Sun** (ez-A) | |
| 8 | **H1 (7)** | **Cell-25** | **Moon** | **Moon** (ez-B) | Full Moon at Node-B ☊ |
| 9 | [Θ1] (8) (BPI) | Cell-30 | Non preserved cells | Sun (ez-B) | |
| 10 | [I1] (9) | Cell-31 | | Moon (ez-A) | Full Moon close to ecliptic limit |
| 11 | [K1] (10) (BPI) | Cell-36 | | Sun (ez-A) | |
| 12, 13 | [Λ1] (11) | Cell-42 | | Moon (ez-A) Sun (ez-B) | New Moon close to Node-B ☊ |
| 14, 15 | [M1] (12) | Cell-48 | | Moon (ez-B) Sun (ez-A) | New Moon close to Node-A ☋ |
| 16, 17 | [N1] (13) (BPI) | Cell-54 | | Moon (ez-A) Sun (ez-B) | New Moon just right on ecliptic limit |
| 18 | [Ξ1] (14) | Cell-59 | | Sun (ez-A) | |
| 19 | [O1] (15) | **Cell-60** | **Moon** | **Moon** (ez-B) | |
| 20 | **Π1 (16)** | **Cell-66** | **Moon** | **Moon** (ez-A) | Full Moon close to Node-A ☋ |
| 21 | **P1 (17)** (BPI) | **Cell-71** | **Sun** | **Sun** (ez-A) | |
| 22 | [Σ1] (18) | Cell-72 | Non preserved Cell | Moon (ez-B) | Full Moon close to Node-B ☊ |
| 23 | **T1 (19)** (BPI) | **Cell-77** | **Sun** | **Sun** (ez-B) | |
| 24 | **Y1 (20)** | **Cell-78** | **Moon** | **Moon** (ez-A) | Full Moon just right on ecliptic limit |
| 25 | [Φ1] (21) (BPI) | Cell-83 | | Sun (ez-A) | |
| 26, 27 | [X1] (22) | Cell-89 | | Moon (ez-A) Sun (ez-B) | New Moon too close to Node-B ☊ |
| 28, 29 | [Ψ1] (23) | Cell-95 | Non preserved cells | Moon (ez-B)<br>Sun (ez-A) | |
| 30 | *[Ω1 cap.] (24). **No relation to cursive ω** | Cell-101 | | Moon (ez-A) | *New Moon far out of ecliptic limit (≥ 5°), No solar eclipse event |
| 31 | [A2] (25) | Cell-106 | | Sun (ez-A) | |
| 32 | [B2] (26) | Cell-107 | | Moon (ez-B) | |
| 33 <u>**Half Saros Cycle**</u> | **Γ2 (27)** | **Cell-113** ΜΕΓΙΣΤΗ or ΤΕΛΕΙΑ ΣΕΛΗΝΗΣ ΕΓΛΕΙΨΙΣ | **Moon** <u>**New Sar period begins**</u> | **Moon** (ez-A)<br>*Shortest Total Lunar eclipse* | **Middle of Cell-113 Full Moon exactly at *Node-A* ☋ and at *Perigee*** |
| 34 | **Δ2 (28)** (BPI) | **Cell-118** | **Sun** | **Sun** (ez-A) | |
| 35 | **E2 (29)** | **Cell-119** | **Moon** | **Moon** (ez-B) | |
| 36, 37 | **Z2 (30)** (BPI) | **Cell-124** | **Moon, Sun** | **Moon** (ez-A)<br>**Sun** (ez-B) | Full Moon on ecliptic limit see **Figure A8** |
| 38, 39 | **H2 (31)** (BPI) | **Cell-130** | **Moon, Sun** | **(–)** *Missing event* Only **Sun** (ez-A) | Full Moon out of ecliptic zone (ez-B), gearing error |
| 40, 41 | **Θ2 (32)** (BPI) | **Cell-136** | **Moon, Sun** | **Moon** (ez-A)<br>**Sun** (ez-B) | New Moon at Node-B ☊ |
| 42, 43 | [I2] (33) | Cell-142 | Non preserved cells | Moon (ez-B)<br>Sun (ez-A) | New Moon close to ecliptic limit |
| 44 | [K2] (34) | Cell-148 | | Moon (ez-A) | |
| 45 | [Λ2] (35) (BPI) | Cell-153 | | Sun (ez-A) | |
| 46 | [M2] (36) | Cell-154 | | Moon (ez-B) | |
| 47 | [N2] (37) | Cell-159 | | Sun (ez-B) | |
| 48 | [Ξ2] (38) | Cell-160 | | Moon (ez-A) | Full Moon close to Node-A ☋ |
| 49 | [O2] (39) | Cell-165 | | Sun (ez-A) | |
| – | No index letter | Cell-166 | Event doesn't exists | (Moon) (Additional event) One Sar after Cell-54 | Full Moon at same ecliptic latitude with New Moon of Cell-54 |
| 50, 51 | **Π2 (40)** (BPI) | **Cell-171** | **Moon, Sun** | **Moon** (ez-A)<br>**Sun** (ez-B) | |
| 52, 53 | **P2 (41)** (BPI) | **Cell-177** | **Moon, Sun** | **(–)** *Missing event* (ez-B), Only **Sun** (ez-A) | Full Moon out the ecliptic zone (gearing error). New Moon at Node-A ☋ |
| 54, 55 | **Σ2 (42)** (BPI) | **Cell-183** | **Moon Sun** | **Moon** (ez-A)<br>**Sun** (ez-B) | New Moon at Node-B ☊ |
| 56 | **T2 (43)** | **Cell-189** | **Moon** | **Moon** (ez-B) and (Sun) (*Additional event*) | New Moon just on ecliptic limit. (Indeterminacy or gearing error) |
| 57 | [Y2] (44) | Cell-195 | Non preserved cells | Moon (ez-A) | |
| 58 | [Φ2] (45) (BPI) | Cell-200 | | Sun (ez-A) | |
| 59 | [X2] (46) | Cell-201 | | Moon (ez-B) | Full Moon at Node-B ☊ |
| 60 | [Ψ2] (47) | Cell-206 | | Sun (ez-B) | |
| 61 | *[Ω2 cap.] | Cell-207 | | Moon (ez-A) | |

<table>
<tr><td></td><td>(48). No relation to cursive ω2</td><td></td><td></td><td></td><td></td></tr>
<tr><td>62</td><td>*2 [A3]<br>(49)<br>(BPI)</td><td>Cell-212</td><td></td><td>Sun (ez-A)</td><td></td></tr>
<tr><td>63,<br>64</td><td>*ω [B3] (50)<br>(BPI)</td><td>Cell-218</td><td></td><td>Moon (ez-A)<br>Sun (ez-B)</td><td></td></tr>
<tr><td colspan="6">* According to DracoNod-V2, the New Moon on Cell-101/index letter omega-1 is far away from the Ecliptic limit and therefore no solar eclipse event corresponds to Cell-101. Additionally, there is no solar eclipse event on Cell-207 (omega-2). Therefore, the solar eclipse event of cursive omega ω in BPI cannot be related to Cell-101/index letter omega-1 and Cell-207/index letter omega-2. Consequently, Cell-101/omega-1 must be related to the lost index letter Ω1 (capital), as also Cell-207/omega-2 with Ω2 (capital). The preserved index letter cursive ω (BPI) should be the 49th or 50th index letter (i.e. after the 48th index letter Ω2/Cell-207). The two last cells including events are Cell-212 and Cell-218; taking into account the second additional index letter 2 (hooked A, see Freeth 2014, 2019; Iversen and Jones 2019), it seems that the ancient Craftsman used the first and the last letters of the Greek alphabet “A” and “Ω” in an “artistic form” (hooked Alpha 2 and cursive ω) and he used them for the 49th and 50th index number. In BPI events classification, Freeth 2019 relates the order of the index letters to the distance from the Node, in L.29 (Freeth 2019, fig.1, 2, 6) <u>ω, 2,</u> Π2, Κ1, Ζ(2), Φ1: according to DracoNod-V2, the New Moon on Cell-218 is farther away from the Node than Cell-212, therefore Cell-218 should have the index letter ω and Cell-212 the index letter 2. According to the authors’ opinion the use of the last letter of the Greek alphabet ω leaves no room for the use of other letters and it must be the last event.</td></tr>
</table>

The total of 64 events in [(24 + 24) +2] = [(A1-…- Ω1)+(A2-…- Ω2) + (2+ω)] = 50 cells including events, consist of 33 lunar eclipse events and 31 solar eclipse events. The revised results by *DracoNod-V2* are in agreement to the preserved eclipse events except the following four results (two missing and two additional events), all of which can be readily justified taking into account the pointers’ marginal positions (moon very close to the ecliptic limits) and the effect of the gearing errors (pointers’ positional errors), which can be slightly alter the lunar pointer’s position (inside the ecliptic zone = additional event, or out of the ecliptic limit = missing event):

1) *Missing*: for Cell 130 (H2) *DracoNod-V2* predicts only a solar eclipse, although a solar and a lunar eclipse are engraved on the cell, (according to *DracoNod-V2*, the Full Moon is out of the ecliptic limit).

2) *Additional*: for Cell-166 *DracoNod-V2* predicts a Lunar eclipse, [one Sar before, i.e. Cell-54/N1, *DracoNod* predicts a solar eclipse event exactly on the ecliptic limit; the index letter N1-solar eclipse is preserved on BPI, L.10(9)].

3) *Missing*: for Cell-177 (P2), *DracoNod-V2* predicts only a solar eclipse, although a solar and a lunar eclipse are engraved on the cell (according to *DracoNod-V2*, the Full Moon is out of the ecliptic limit).

4) *Additional*: for Cell 189 (T2) *DracoNod-V2* predicts an additional event (solar eclipse) although only a lunar eclipse is engraved (according to *DracoNod-V2*, the New Moon exactly on the ecliptic limit).

In general, events that are Close-to/On/Out of the Ecliptic Limits present a degree of uncertainty (or indeterminacy) due to the gearing errors. In different functional models of the Mechanism (which have different mechanical errors), these events could be either calculated as events or not. Results that are marginally at the limits of calculation are affected by the construction limits of the instrument.

## 6. Eclipse events classification according to Eclipse Magnitude – Correlating the Eudoxus Papyrus to the BPI of the Antikythera Mechanism

On the Back Plate Inscription (Freeth 2014 and 2019; Anastasiou et al., 2016a; Iversen and Jones 2019), the words Μ]ικραί-Βορείου-Μέσαι-Με[γά]λ<αι[5]>-Νότον-Μέσαι-Νότου-Μικραί (Minor-North-Medium-Large-South-Medium-South-Minor, L.3, 8, 15, 28, 36 in Iversen and Jones 2019) are referred on the eclipse classification text, **Figure 8**.

These words are placed in "mirror symmetry" to the word Με[γά]λ<αι>. Taking into account the words Βορείου (North) and Νότον (South), the two times repeated words Μικραί (Minor) and Μέσαι (Medium) are related to Βορείαι Μικραί/Βορείαι Μέσαι (North of Node) and to Νοτίαι Μέσαι/Νοτίαι Μικραί (South of Node).

Contrary to all the other words related to eclipse magnitude characterization, the word Μεγάλαι (Large) is referred only once and it corresponds to the Northern Large eclipses[6]. Note that there exists a free/empty space for additional engraved inscriptions, if the ancient Craftsman wanted to repeat for a second time the description for the Νοτίαι Μεγάλαι Εγλείψεις/Souther Large eclipses. Thus the engraved word Μεγάλαι correspond to:

- either independently North/South of Node (this is not in accordance to the word ΘΡΑΙΚΙΑΣ in L.11, which is a NW or NNW wind),
- either the ancient Craftsman did not include the Νοτίαι Μεγάλαι (Southern Large) on the BPI, because they were not detected Νοτίαι Μεγάλαι eclipses according to the Draconic gearing.

After careful examination of the eclipse events' Draconic phase (calculated by *DracoNod-V2* program), no eclipse events that could be classified as Νοτίαι Μεγάλαι – Southern Large eclipses were detected, and so this explains reasonably well why the word Μεγάλαι (Νοτίαι eclipses) were not referred for a second time on the BPI[7].

Today, part of the solar eclipse classification in BPI and all of the text for the lunar eclipse classification are lost.[8] The lost inscriptions for the solar eclipse classification must have been engraved on the right blank area between the Metonic and Saros spirals, see **Figures 8** and **11**.

[5] In BPI is engraved (by mistake) με]γάλην (singular) instead of με]γάλαι (plural), see Iversen and Jones 2019, p. 480-481.

[6] In Iversen and Jones 2019, p. 476: Ἀπό Θραικί<ου> πε[ρι]ίστανται δέ κ[αί] καταλήγο[υσι] προς Ἀπηλ[ι]ώτην, μεγάλ<αι>… From the Northwest they wheel about and also subside to the east; Large… i.e. the shadow becomes visible from Northwest and ends to the East: this direction refers to Northern (above Ecliptic) large obscuration eclipses.

[7] The solar eclipse event on Cell-218/Ω is the first event for Sector (Νοτίαι) Μέσαι – (Southern) Medium (the Draconic pointer is located at the limit between Southern Large and Southern Medium Sectors). The solar eclipse event on Cell-42/Λ1 is close to this limit.

[8] Anastasiou et al., 2016a suggested that the right side of the Saros Dial was dedicated to all of the solar eclipse events and the left one for all the lunar eclipse events. A different approach and a different arrangement of the Back Plate Inscriptions is presented in Freeth 2014 (Fig. 9) and 2019 (Fig. 4).

In Eudoxus papyrus (or *Ars Eudoxi*, in P. Par. 1, Ed. Blass 1887) written in 165/164 BC (see Jones 1999), in col. XVIII, 10-15 and col. XIX, 1-5 is mentioned that when the Moon, the Node and the Sun are all on a straight line, then the **Μεγίστη ἡλίου ἔγλειψις**[9] occurs (**The Greatest** solar eclipse occurs exactly at the Node i.e. the centers of the two celestial bodies coincides). The word **Μεγίστη**[10] in singular, i.e. it is the only one greatest eclipse. When the Moon is ἀποτέρω καί ἀποτέρω[11] (on either side of the Node)[12], the solar eclipses are ἐλάττους καί ἐλάττους (lesser eclipses occur away and on either side of Node).

The word **Μείζους/(Μείζονες**, Major in plural) is also referred (col. XIX, 11-15): Αἱ δέ μείζους ἁψιδοειδεῖς (the major solar/lunar eclipses appear arch-shaped, like a bridge), αἱ δέ [ἔ]τ[ι] μείζους ὠιοειδεῖς (the major lunar eclipses appear oval-shaped).

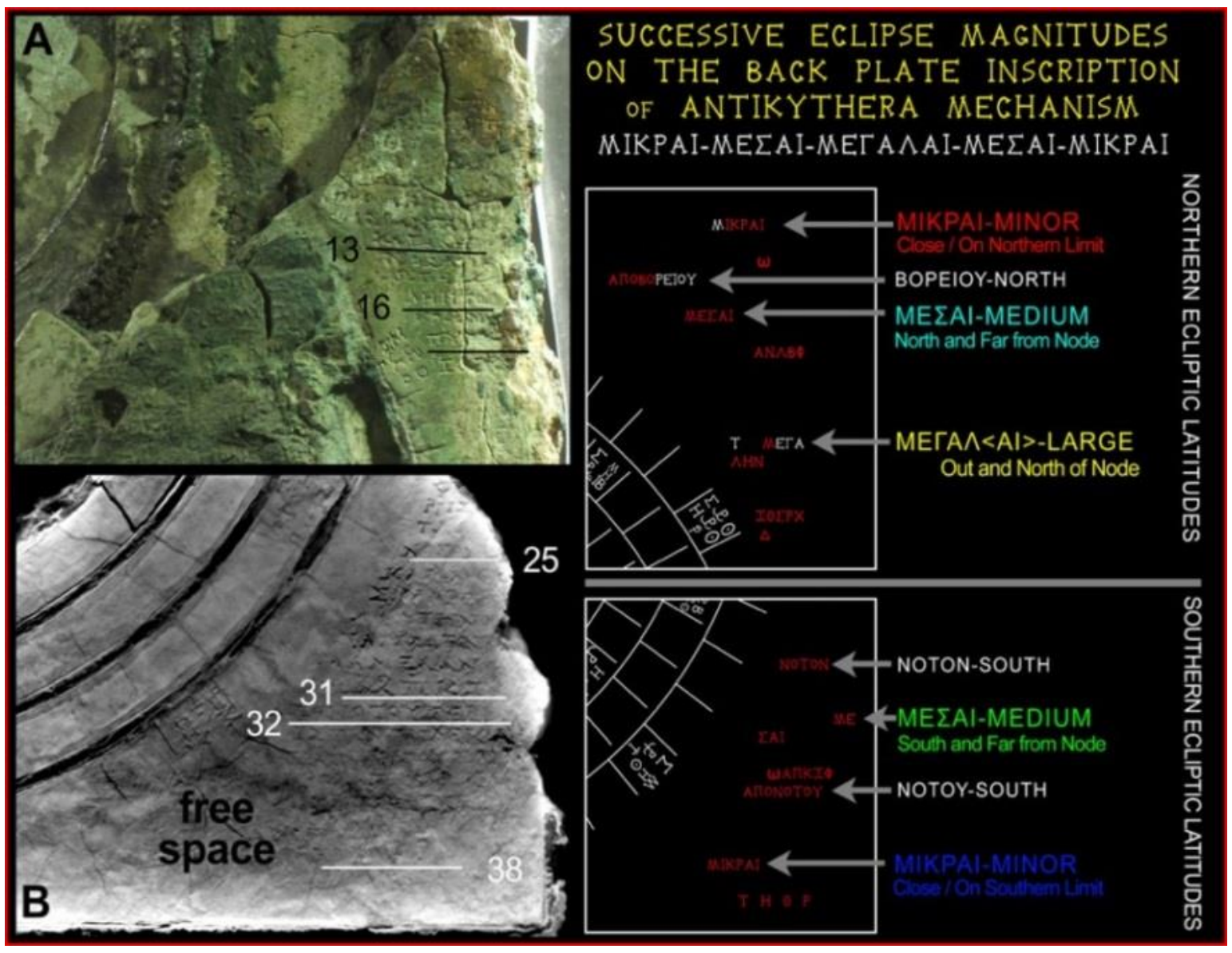

**Figure 8:** A) Close-up on the Back Plate Inscription visual photograph of Fragment A2 (photo by first author). B) Multi-combined AMRP X-ray tomography of Fragment F1 at the area with the preserved Inscriptions. The tomography slices are oriented to the surface with inscriptions. The orientation and the slices were processed by the authors using Real 3D VolViCon software. The numbers correspond to the numbering of lines according to Iversen and Jones 2019. Note that a free/empty space exists to the left of lines 32-38. Right panel, the preserved words μικραί, βορείου, μέσαι, με[γά]λ<αι>, νότον, μέσαι, νότου, μικραί are presented according to their corresponding ordered position on the Back plate with red (preserved) and white letters (see small differences on Freeth 2014 and 2019; Anastasiou 2016; Iversen and Jones 2019; Jones 2020). The order of these words has a mirror symmetry relative to the word με[γά]λ<αι>, which is engraved only once.

Moreover, the text differentiates the **Μείζους**/Major eclipses from the **Ἐλάσσους**/lesser eclipses: Αἱ μέν ἐλάσσους μηνοειδεῖς, αἱ δέ μείζους ἁψιδοειδεῖς (the lesser eclipses appear meniscus-shaped, while the major eclipses appear arch-shaped). The difference between the meniscus and the arch-shaped configuration is focused on the thickness of the

9 See LSJ. The word ἔγλειψις is referred in Eudoxus papyrus, instead of the usual ἔκλειψις (Blass 1887, col. XVIII – XX; see also Papathanassiou 2010), as also ἐγ δεξιῶν instead of ἐκ δεξιῶν (Blass 1887, col. I, 10). On the Antikythera Mechanism the word ἐγλειπτικοὶ is written, instead of ἐκλειπτικοὶ (Bitsakis and Jones 2016b, p.235).

10 Simplicius (II, 12, p.505) writes the word ΠΑΝΤΕΛΕΙΣ ΕΚΛΕΙΨΕΙΣ (full coverage)

11 In this text the word ἀποτέρω refers to eclipses out of Node/off-axis geometry, in a band area before/after a Node.

12 See also Neugebauer 1975, p.686-689.

central area: ἁψιδοειδεῖς (arch-shaped eclipses) have a very thin central area, like an arch (see **Figures A10 and A11**), meaning that Major eclipses have a very high eclipse magnitude/obscuration.

The word Μεγίστη[13] is directly related to the highest Eclipse Magnitude/Eclipse Obscuration[14] (the centers of the Moon and the Sun coincide, max or full coverage), the word ἐλάττους with the lower/low/small Eclipse Magnitude/Obscuration) and the word Μείζους with eclipses of a bit less obscuration than Μεγίστη obscuration. The eclipse magnitude/obscuration according to Eudoxus papyrus follows the order: Μεγίστη, Μείζους, Ἐλάττους καί Ἐλάττους (*before Node/North Far from Node* and *after Node/South Far from Node*), see **Table 4**.

Correlating the preserved words from Antikythera Mechanism BPI (μικραί, βορείου, μέσαι, με[γά]λ<αι>, νότον, μέσαι, νότου, μικραί) and the words/information from Eudoxus papyrus (Μεγίστη Ἔγλειψις and Μείζους/Μείζονες) and applying the ancient Greek Grammar, these words can describe all kinds of eclipse magnitudes, i.e. the Moon's position relative to the Node or to an Ecliptic limit[15], see **Table 4**.

**Table 4:** Correlating the Eclipse Magnitude/Obscuration nomenclature according to Eudoxus papyrus (Blass 1887) and the BPI of the Mechanism (Freeth 2014 and 2019; Anastasiou 2016; Iversen and Jones 2019). The two nomenclatures present mirror symmetry (see **Figures A9** and **A10** in **Appendix-A**).

<table>
<tr><th colspan="2">Eudoxus papyrus nomenclature and Description</th><th colspan="2">Antikythera Mechanism<br>Back Plate Inscription</th></tr>
<tr><td rowspan="3">Ἐλάττους<br>Minor/lesser<br>(plural)</td><td rowspan="3">(Before) Node/<br>Outside and Far of Node<br>(northern ecliptic latitudes)</td><td>Μ]ικραί<br>(L.3)</td><td>Βορείαι Μικραί ἐγλείψεις<br>Minor Northern eclipses</td></tr>
<tr><td>Μέσαι<br>(L.8)</td><td>Βορείαι Μέσαι ἐγλείψεις<br>Medium Northern eclipses</td></tr>
<tr><td rowspan="3">Με[γά]λ<αι><br>(L.15/16)<br>Lost inscriptions several Lines before L.1 (…………….)</td><td rowspan="3">Βορείαι Μεγάλαι ἐγλείψεις<br>Northern Large Eclipses<br><br>(………………………………..)</td></tr>
<tr><td>Μεγίστη<br>The Greatest<br>(only one)<br>*****<br>Μείζους<br>Major<br>(plural)</td><td>Exactly<br>at the Node<br><br>*****<br>Close to the Node<br>(before/after)</td></tr>
<tr><td rowspan="3">Ἐλάττους<br>Minor/lesser<br>(plural)</td><td rowspan="3">(After) Node/<br>Outside and Far of Node<br>(southern ecliptic latitudes)</td></tr>
<tr><td>Μέσαι<br>(L.28/29)</td><td>Νοτίαι Μέσαι ἐγλείψεις<br>Medium Southern eclipses</td></tr>
<tr><td>Μικραί<br>(L.36)</td><td>Νοτίαι Μικραί ἐγλείψεις<br>Minor Southern eclipses</td></tr>
</table>

[13] Μεγίστη (greatest-superlative), Μείζων (greater-comparative), Μεγάλη (great-positive), see μέγας in LSJ. The word μεγάλαι (plural), include eclipses close and on either side of the Node (but out of Node).

[14] The magnitude of a solar eclipse is the fraction of the Sun's diameter that is covered by the Moon and the magnitude of lunar eclipse is the fraction of the Moon's diameter covered by the Earth's umbra. See an explanatory interactive app in GeoGebra https://www.geogebra.org/m/SnZ7QGTJ. The Eclipse obscuration is the fraction of the Sun's surface area covered by the Moon. Generally, an eclipse magnitude at 50% corresponds into 40% obscuration of the solar disc, at 75% into 68.5% obscuration, and at 25% into 14.5%.

[15] Cleomedes (Ziegler 1891, p.214) classifies the eclipses as τελείαν (Perfect) ἔκλειψιν ποιεῖται καὶ ὑψηλοτάτη καὶ προσγειοτάτη or ταπεινοτάτη καὶ μέσως ἔχουσα (highest, lowest and mean eclipse).

The words Μεγίστη and Μείζονες are missing from the BPI of the Mechanism: The most probable scenario is the unique reference of the (only one) greatest solar eclipse per Saros (event A1-Μεγίστη Ἔγλειψις) and the (only one) greatest lunar eclipse per Saros (event Γ2-Τελεία or Μεγίστη Ἔγλειψις), both taking place exactly at a Node. This reference could have existed at the beginning of the right BPI column (solar eclipse events) and on the left BPI column (lunar eclipse events, today lost), see **Figure 11**. The inclination of the ecliptic (as it projected on the sky) relative to the Equator it continuously varies through the time and the date by ±23.5°. The Moon actually moves from west to east [in max inclination ±(23.5°± 5.14°) relative to the Equator). During the Greatest eclipse the centers of the Moon and the Sun are perfectly (or almost perfectly) aligned, a probable engraved description for the Greatest eclipse (exactly at the Node) and Major eclipses (close/about to the Node) could have been: Ἀπό Ζεφύρου ἄρχονται δε καί καταλήγουσιν πρός Ἀπηλιώτην, και Μεγίστη καί Μείζονες, το δε χρώμα μέλαν, They begin from Zephyrus (west) and they end on Apeliotes (east), The Greatest (solar) eclipse and the Major (eclipses), the color is black or Ἡλίου Μεγίστη ἔγλειψις και Μείζονες, ἀπό Ζεφύρου ἄρχονται δε καί καταλήγουσιν πρός Ἀπηλιώτην, το δε χρώμα μέλαν, The Greatest eclipse of the Sun and the Major (eclipses), they begin from Zephyrus and they end on Apeliotes, the color is black.

Due to the inclination of the ecliptic relative to the Equator the direction of the Greatest eclipse could also be from Argestes to Eurus or from Lips to Caecias.

By applying the initial starting date for the Saros and the Metonic spirals of the Mechanism on 22$^{th}$/23$^{rd}$ December 178 BC (Voulgaris et al., 2023a) and the time of the Greatest solar eclipse around to 1$^{st}$ daily hour i.e. at the eastern horizon, the Ecliptic and the Equator (direction) are parallel between them, and therefore the orbital direction of the Moon is west-east (Zephyrus-Apeliotes).

*Ptolemy* IV.6 (Heiberg 1898, p.302) uses the word **Τελεία** for a total lunar eclipse when the Moon is completely and symmetrically covered by the Earth's shadow (i.e. the Moon passes exactly or very close to the from umbra's center, therefore the Moon lies exactly or very close to the Node). When the Moon lies close to umbra's center (deep total lunar eclipse) the whole lunar disc turns red or beige. If the total lunar eclipse is not well centered, the parts that are closer to the umbra limits are brighter than the rest of the lunar disc and a bright arc appears on the lunar disc during totality.

*Cleomedes* II.4 (Ziegler 1891) writes about the Τελεία (ἡλίου) ἔκλειψις: … ἐν ταῖς τελείαις τῶν ἐκλείψεων, ὅτε ἐπὶ μιᾶς εὐθείας γίνεται τά κέντρα τῶν θεῶν (the Perfect solar eclipses occur when the ***Centers of Gods*** are aligned in a straight line), **Figure 9**. He also mentions (II.6) the words **Τελείαι** for total and **Μερικαί** for partial lunar eclipses.

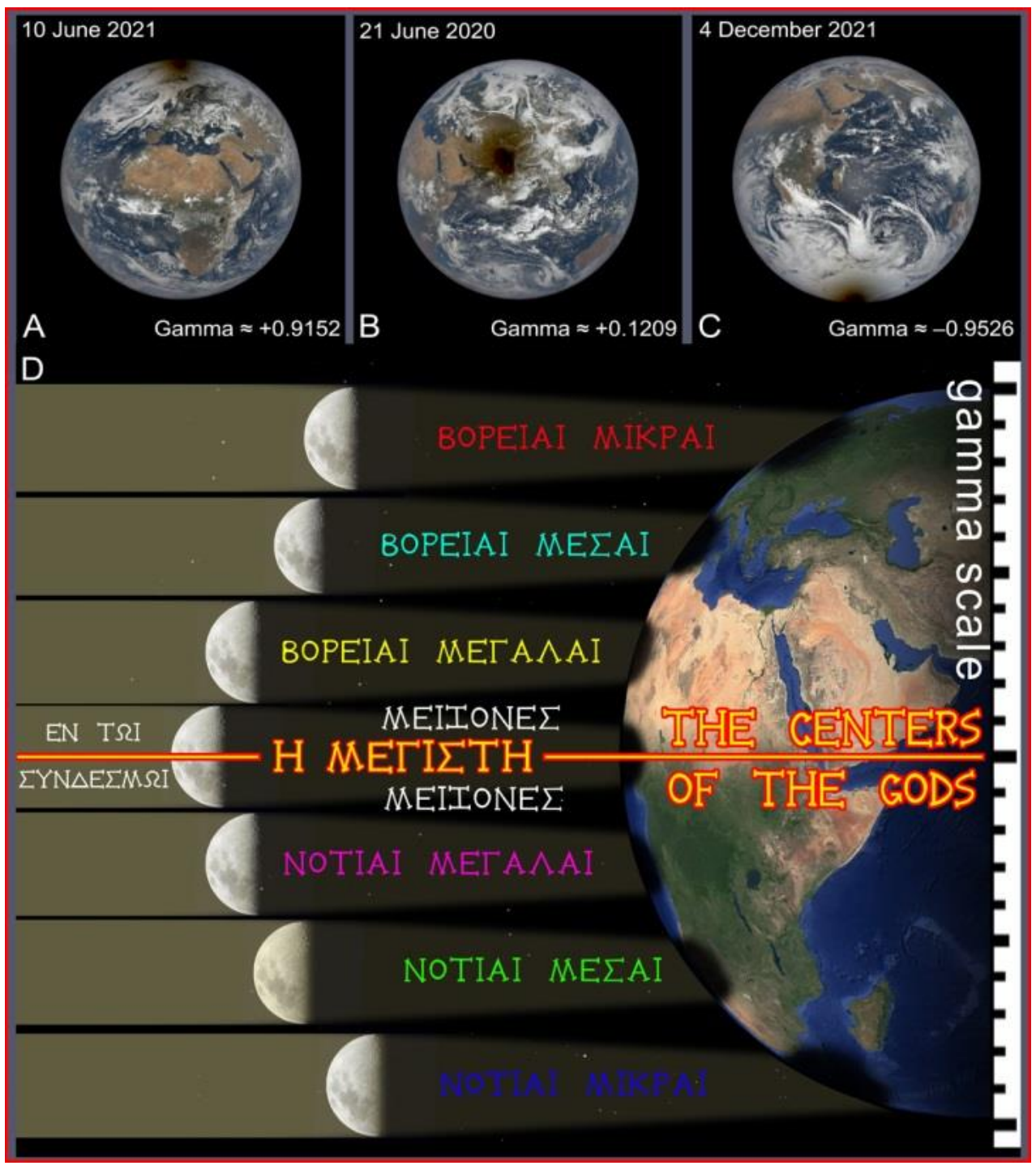

**Figure 9: A**, **B**, **C**, The Moon's shadow was recorded by the EPIC System of the DSCOVR Satellite by NASA from 1 million miles away at three different dates during the solar eclipse of: 10th June 2021 (shadow around to North Pole), 21th June 2020 (central eclipse occurs at Tropic of Cancer) and 4th December 2021 (shadow around South Pole, see also **Figure A-9**). The position of the Lunar shadow as it crosses the Earth's surface depends on the distance of the Moon from the Node (which defines the eclipse Gamma, Meeus 1991 and 1997; NASA Eclipse Web Site[16]) and the Seasons, as they define the Declination of the Sun, **D)** The profiles of the lunar shadows (in different eclipse Gamma and not on scale), according to the new results of the Antikythera Mechanism eclipse classification by the present work. The red/yellow line depicts the *Centers of the Gods* according to Cleomedes' definition for the alignment on the straight line of the Sun-Moon-Earth's centers.

## 7. Ecliptic Zone division in Sectors according to the eclipse events classification

Based on the integrated nomenclature (seven words) describing the eclipse magnitude(s), a correlation with specific areas on each ecliptic zone can be defined. Each of the two Ecliptic Zones A and B is divided in a thin linear area of Node named by the authors Ἐγλείψεις ἐπὶ τῆς τοῦ Συνδέσμου Χώρας (eclipses On/About-on the Area of Node) and in six unequal Sectors. The area with the phrase Ἐγλείψεις ἐπὶ τῆς τοῦ Συνδέσμου Χώρας refers to:

i) The **Maximum eclipse** (the only one) **Μεγίστη [0]** that occurs exactly at the Node, and

ii) **Major eclipses** (more than one) **Μείζονες** (or **Μείζους**) **[+ 0 -]**, that occur *Very Close to the Node* (just before/just after the Node - either North/South of the Node), inside the Area of Node.

The six Sectors consist of:

- the **Sector-NL/Μικραί (Βορείαι) ἐγλείψεις**: *Close to/On the Northern Ecliptic Limit*,
- the **Sector-NF/Μέσαι (Βορείαι) ἐγλείψεις**: N*orth and Far from the Area of the Node*,
-the **Sector-NCA/Μεγάλαι (Βορείαι) ἐγλείψεις**: *North and outside of the Area of the Node,*
- the **SCA/Μεγάλαι (Νοτίαι) ἐγλείψεις**: *South and outside of the Area of the Node*,

[16] https://eclipse.gsfc.nasa.gov/SEcat5/catalog.html, see also the representative graphics in *Variations in Gamma*: https://freehostspace.firstcloudit.com/steveholmes/saros/gamma1.htm

-the **Sector-SF/Μέσαι (Νοτίαι) ἐγλείψεις**: *South and Far from the Area of the Node*,
- the **Sector-SL/Μικραί (Νοτίαι) ἐγλείψεις**: *Close to/On the Southern Ecliptic Limit*,

see **Table 5** and **Figure 10**. High magnitude eclipses present a high probability for detectability/visibility (also dependent on the season they occur).

**Table 5:** An integrated nomenclature for the solar and lunar Eclipse Magnitude/Obscuration based on the Antikythera Mechanism BPI and on Eudoxus papyrus. The pattern begins with the most important eclipses using the words Μεγίστη-The Greatest (maximum obscuration) and Μείζονες- Major (almost maximum obscuration) and then it continues with the Northern parts of the ecliptic zone, and ends with the Southern parts. The most important eclipse Cell-1/A1, Η ΤΟΥ ΗΛΙΟΥ ΜΕΓΙΣΤΗ ΕΓΛΕΙΨΙΣ should be noted as a special reference-heading of the BPI right column and in the same manner Η ΤΗΣ ΣΕΛΗΝΗΣ ΤΕΛΕΙΑ (or ΜΕΓΙΣΤΗ) ΕΓΛΕΙΨΙΣ Cell-113/Γ2, the heading of the BPI left column. Based on this nomenclature, each ecliptic Zone (max epicenter angle ≈29°) is divided in six (five)[17] unequal Sectors and a linear central area around the Line of Nodes, see **Figure 10**. The width (degrees) for each Sector depends on the defined percentage range of coverage (see last column). Note that the Μεγάλαι/Large eclipses have a large percentage range of about 30%.

| Sectors of Ecliptic Zone | BPI Nomenclature and ordered position for the Eclipse Magnitude | Moon position relative to Node/Ecliptic limit | Sectors of Ecliptic Zone | Eclipse magnitude[18] |
|---|---|---|---|---|
| **Node** | **Ἡ τοῦ Ἡλίου Μεγίστη Ἔγλειψις** The Greatest Eclipse | Exactly at the Node | **Line-0** | **100% (≥100 total) (93-99.9% annular)** |
| **Area of the Node** | ******** **Μείζονες ἐγλείψεις** Major Eclipses | ******** Very Close to the Node and on either side, regardless of North/South of Node | ******** **Linear area [+ 0 -]** | ******* **≈ 90–99% (range ≈9%)** |
| **Northern Ecliptic latitudes** | **Μεγάλαι Βορείαι ἐγλείψεις** Large eclipses | Outside and North of the Area of Node Only Northern Large eclipses | **Sector NCA** (central band area of the ecliptic zone) | **≈ 60–90% (range ≈30%)** |
| | **Μέσαι Βορείαι ἐγλείψεις** Medium Northern eclipses | (only) North and Far from the Area of the Node | **Sector FN** (area) | **≈ 40–50–60% (range ≈20%)** |
| | **Μικραί Βορείαι ἐγλείψεις** Minor Northern eclipses | Close to/On the Northern Ecliptic limit | **Sector NL** (area) | **≈ 10–40% (range ≈30%)** |
| **Southern Ecliptic latitudes** | **Μέσαι Νοτίαι ἐγλείψεις** Medium Southern eclipses | (only) South and Far from the Area of the Node | **Sector FS** (area) | **≈ 40–50–60% (range ≈20%)** |
| | **Μικραί Νοτίαι ἐγλείψεις** Minor Southern eclipses | Close to/On the Southern Ecliptic limit | **Sector SL** (area) | **≈ 10–40% (range ≈30%)** |

The division of the eclipse zone is calibrated as close as possible to the preserved index letters classification, but as the gearing errors alter/("*deform*") the pointers' theoretical position, they create some declinations or mismatches, see **Figure 10** and **Figures A4-A7** in **Appendix-A**). The eclipse events classification is presented on **Table 6**, assuming a perfect instrument without mechanical errors.

[17] There isn't any Sector/text in BPI for the Southern Large eclipses, (probably) because such eclipse events were not detected (according to *DracoNod* program), therefore no reason existed to refer/engraved them.
[18] See the interactive app in GeoGebra https://www.geogebra.org/m/SnZ7QGTJ.

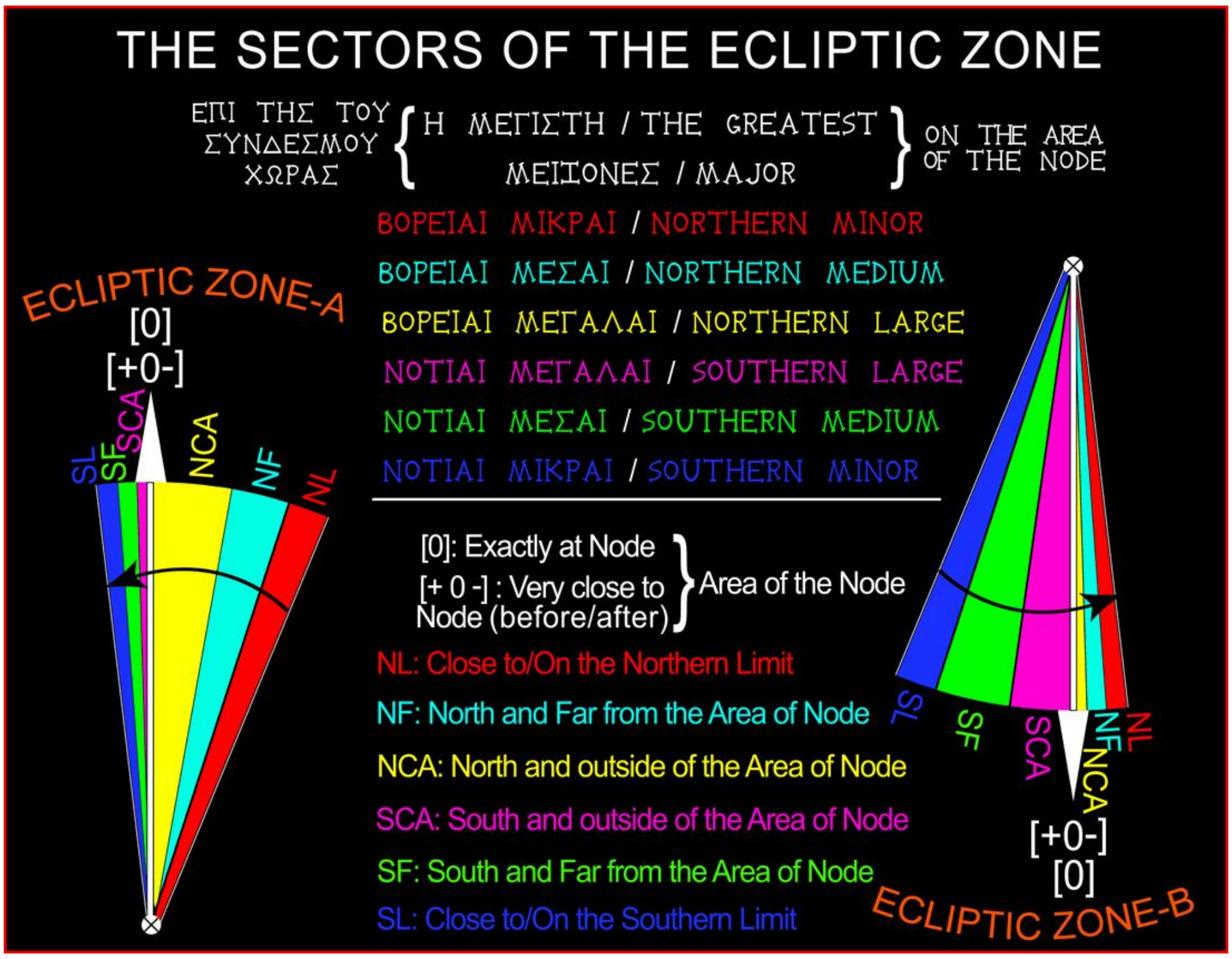

**Figure 10:** The Ecliptic Zones A and B are divided into the (linear band) Area of Node and in six unequal Sectors. The Μεγίστη (The Greatest) and Μείζονες (Major) eclipses occur on/close to the linear band of the Node (white line). The southern parts of the Ecliptic Zone-A/☋ are suppressed as also the northern parts of the Ecliptic Zone-B/☊. Note that the boundaries for each of the six ecliptic Sectors cannot be precisely defined because the gearing errors alter the Draconic pointer position, thus changing the events' classification, see **Figures A4-A7** in **Appendix-A**.

## 8. Epilogue

Based on the preserved data of the Antikythera Mechanism (the shaft-a: an output with unknown operation and the unplaced gear-r1 of Fragment D) and correlating them with the observation that the important Draconic cycle is not represented on the Mechanism, a hypothesis was created. This hypothesis (see *Protocol of Mandatory Parameters and Scientific Methodology for a Research Quality Bronze Reconstruction of the Antikythera Mechanism in* Voulgaris et al., 2025a, Appendix p.277), creates a number of consequences, impacts, results and data (Feynman 1964 and 1967). Many questions are answered, several results are well explained, and the eclipse events' specific sequence engraved on the Saros cells are justified **Figure 11**.

The description for the Draconic circle, pointer and scale could also have been presented in the *Instruction Manual* of the Mechanism (Back Cover Inscription, see Bitsakis and Jones 2016b), as ΣΕΛΗΝΗΣ ΑΠΟΚΑΤΑΣΤΑΣΙΣ ΚΑΤΑ ΠΛΑΤΟΣ, but a large percentage of the text, about 34 Lines (≈38% of the BCI) is not preserved today.

**Table 6:** The theoretical solar and lunar eclipse events' classification (assuming no gearing errors) calculated according to the 5+2 Sectors of the ecliptic zone(s) based on *DracoNod-V2*. Colored events are according to the (well) preserved index letters on Back plate inscription (Freeth 2014 and 2019; Anastasiou 2016; Iversen and Jones 2019; Jones 2020). A special reference for Μεγίστη solar eclipse and a special reference for Τελεία lunar eclipse are added. The deviation(s) between the theoretical and the existing engraved events' classification can be attributed to the gearing errors, the gear teeth

non-uniformity and the "*deformed*" ecliptic limits due to the gearing errors. Note that the ecliptic zone Sectors are not in equal angular dimensions. The questionable evens (?) located just between two Sectors.

| Solar eclipse classification | | | | | | | |
|---|---|---|---|---|---|---|---|
| **BPI** | **ΒΟΡΕΙΑΙ ΜΙΚΡΑΙ** | **Α2, Ν1, Λ2, Β1, Φ2 ΒΟΡΕΙΑΙ ΜΕΣΑΙ** | **Ζ1, Θ2, Σ2, Ρ1, Χ1, Δ2 ΜΕΓΑΛΑΙ** | **MISSING/LOST HEADING Right BPI (solar) Left BPI (lunar)** | **Non engraved in BPI** | **ω, ϩ, Π2, Κ1, Ζ2, Φ1 ΝΟΤΙΑΙ ΜΕΣΑΙ** | **Τ1, Η2, Θ1, Ρ2 ΝΟΤΙΑΙ ΜΙΚΡΑΙ** |
| **New Moon on Ecliptic zone Sector** | **<u>Sector NL</u> Close to/On Northern Ecliptic Limit ΒΟΡΕΙΑΙ ΜΙΚΡΑΙ Northern Minor** | **<u>Sector NF</u> North & Far from the Node ΒΟΡΕΙΑΙ ΜΕΣΑΙ Northern Medium** | **<u>Sector NCA</u> North & Outside of the Area of Node ΒΟΡΕΙΑΙ ΜΕΓΑΛΑΙ Northern Large** | **<u>Area of Node</u> [0] Η ΤΟΥ ΗΛΙΟΥ ΜΕΓΙΣΤΗ ΕΓΛΕΙΨΙΣ The Greatest New Moon Exactly at Node**<br>**ΜΕΙΖΟΝΕΣ [+ 0 -] ΕΓΛΕΙΨΕΙΣ/Major New Moon before/after Node** | **<u>Sector SCA</u> South and Outside of the Area of Node ΝΟΤΙΑΙ ΜΕΓΑΛΑΙ Southern Large** | **<u>Sector SF</u> South & Far from the Node ΝΟΤΙΑΙ ΜΕΣΑΙ Southern Medium** | **<u>Sector SL</u> Close to/ On the Southern Ecliptic limit ΝΟΤΙΑΙ ΜΙΚΡΑΙ Southern Minor** |
| | **12-Γ1**,<br>54-Ν1,<br>59-Ξ1 | **7-Β1**,<br>106-Α2,<br>153-Λ2,<br>200-Φ2 | **24-Ζ1**,<br>36-Κ1,<br>**71-Ρ1**,<br>83-Φ1,<br>89-Χ1 (?),<br>**118-Δ2**,<br>,<br>**136-Θ2**,<br>165-Ο2,<br>**183-Σ2,**<br>212-ϩ**(A3)** | **THE GREATEST**<br>1-Α1 ☋<br>**MAJOR**<br>48-Μ1 (?),<br>**177-Ρ2** | 42-Λ1 (?),<br>48-Μ1 (?),<br><br>**Probably No Solar eclipse events occurred** | 42-Λ1 (?),<br>95-Ψ1,<br>**124-Ζ2**,<br>142-Ι2,<br>**171-Π2**,<br>218-ω(B3) | ,<br>,<br>159-Ν2,<br>206-Ψ2 |
| **Lunar eclipse classification (lost)** | | | | | | | |
| **Full Moon on Ecliptic zone Sector** | **<u>Sector NL</u> Close to/On Northern Ecliptic Limit ΒΟΡΕΙΑΙ ΜΙΚΡΑΙ Northern Minor** | **<u>Sector NF</u> North & Far from Node ΒΟΡΕΙΑΙ ΜΕΣΑΙ Northern Medium** | **<u>Sector NCA</u> North & Outside of the Area of Node ΒΟΡΕΙΑΙ ΜΕΓΑΛΑΙ Northern Large** | **<u>Area of Node</u> [0] Η ΤΗΣ ΣΕΛΗΝΗΣ ΜΕΓΙΣΤΗ/ΤΕΛΕΙΑ ΕΓΛΕΙΨΙΣ The Greatest Full Moon Exactly at the Node**<br>**ΜΕΙΖΟΝΕΣ [+ 0 -] ΕΓΛΕΙΨΕΙΣ/Major Full Moon before/after Node** | **<u>Sector SCA</u> South & Outside of the Area of Node ΝΟΤΙΑΙ ΜΕΓΑΛΑΙ Southern Large** | **<u>Sector SF</u> South & Far from the Node ΝΟΤΙΑΙ ΜΕΣΑΙ Southern Medium** | **<u>Sector SL</u> Close to/ On the Southern Ecliptic limit ΝΟΤΙΑΙ ΜΙΚΡΑΙ Southern Minor** |
| | **119-Ε2**,<br>**124-Ζ2**,<br>**171-Π2**,<br>218-ω(B3) | **7-Β1**,<br>42-Λ1,<br>89-Χ1,<br>**136-Θ2**, | **19-Ε1**,<br>54-Ν1,<br>72-Σ1,<br>101-Ω1,<br>148-Κ2,<br>**183-Σ2**,<br>195-Υ2 | **THE GREATEST**<br>**113-Γ2 ☋**<br>**MAJOR**<br>**25-Η1**,<br>**66-Π1**,<br>160-Ξ2 (?),<br>201-Χ2 | 107-Β2(?),<br>154-Μ2,<br>160-Ξ2 (?) | 13-Δ1,<br>**60-Ο1**,<br>107-Β2(?),<br>**189-Τ2**,<br>207-Ω2 | 31-Ι1,<br>48-Μ1,<br>78-Υ1,<br>95-Ψ1,<br>142-Ι2,<br>**177-Ρ2** |

The Draconic pointer offers the second component (ecliptic latitude) for the Moon's position in the sky (and the absence of this cycle creates indeterminacy in the Lunar position): E.g. when the Lunar pointer aims at the 1st subdivision of Leo, where exactly is the Moon located? On, Below, or Above the Ecliptic line? (As Regulus is located on the Ecliptic, if the Moon is On the Ecliptic, this means On, or Close proximity to Regulus). However, without the information of lunar latitude, the Moon could be 2°, 3°, 4°, or 5° away from Regulus, either above or below it (in max positional error of 10°).
The positional correlation of the Lunar pointer (ecliptic longitude) and the Draconic pointer (ecliptic latitude) offers information for a possible occultation of a star by the Moon. A number of star occultations/conjunctions by the Moon recorded by the ancient astronomers, also mentioned in Ptolemy (see Toomer 1984, p. 334-338).

The hypothesis of the Draconic gearing creates a crucial implication for the Antikythera Mechanism: the eclipse events were "*artificially*" created via a pure mechanical procedure, using only the Antikythera Mechanism and without the need of a non-directly related information: The ancient Craftsman created a device to measure time, to detect future astronomical and social events, based on the lunar cycles, as the social life and the religious activities of the ancient Greeks were dependent and regulated by the lunar phases (synodic month, calendars, the start of the athletic Games, household celebrations, sacrifices[19]).
To achieve this task, the ancient Craftsman introduced to his device the representation of the four lunar cycles. At this time he didn't know the exact eclipse events' sequence. After the initial calibration of the Mechanism's pointers, he started to rotate the Input of the Mechanism - the Lunar Disc - by aiming successively to the Golden sphere–New Moon and in the opposite direction–Full Moon. At the same time he checked the position of the Draconic pointer-Lunar ecliptic latitude, at the right side of the Mechanism; if the pointer was inside the ecliptic zone, he engraved the symbol H (solar eclipse) or Σ (lunar eclipse) (Freeth et al., 2006 and 2008) on the cell that the Saros pointer aimed, as also the number (index letter) of the cell with event(s) and the events' predicted hour (Voulgaris et al., 2023b, p.22-25). Additionally, he classified each event according to the specific position of the Draconic pointer inside the ecliptic zone. If the Draconic pointer was out of the ecliptic zone, the Saros cell remained blank.

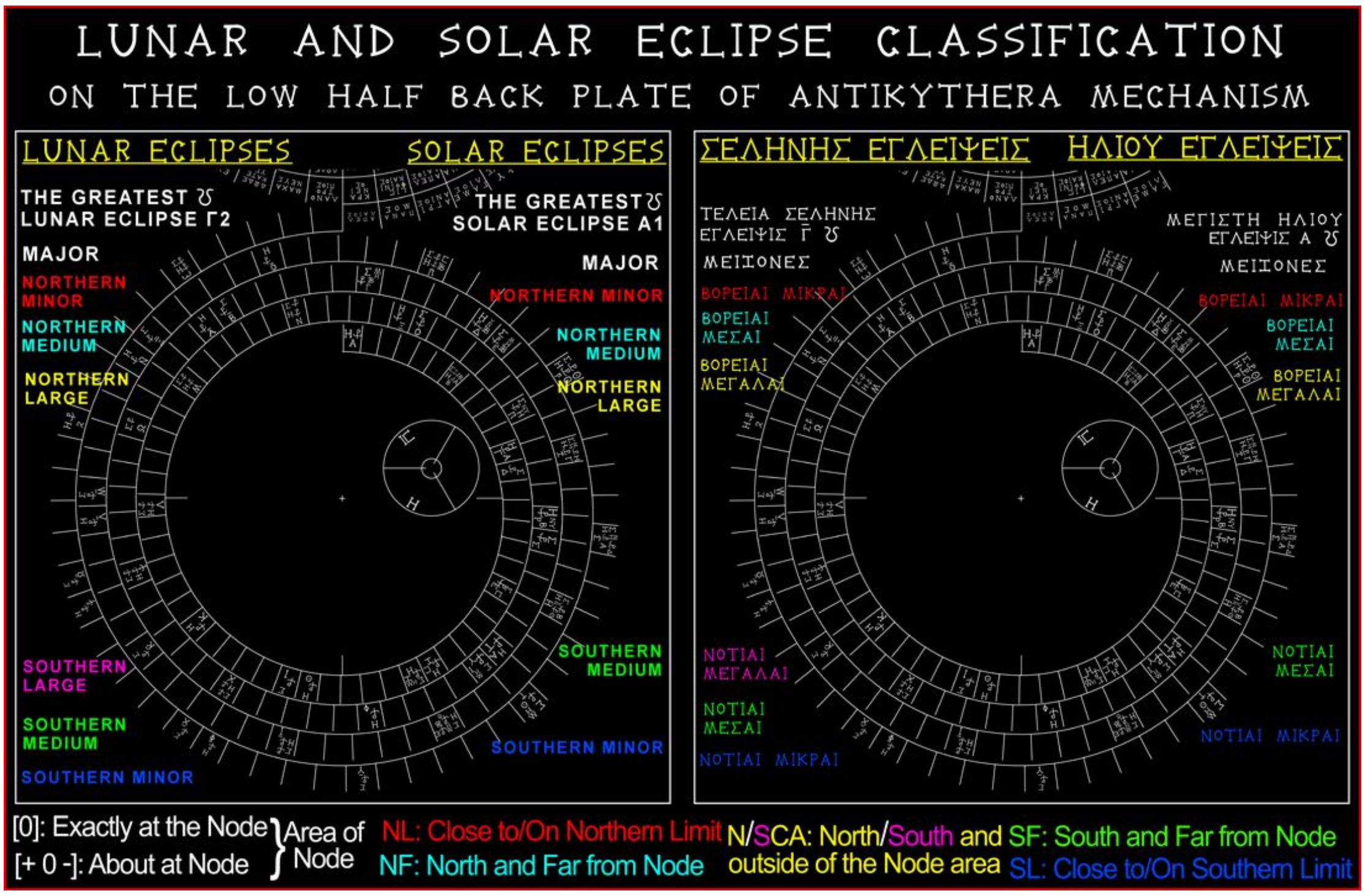

[19] E.g. Hekate's Deipnon at the end of lunar month and Noumenia at the new month begin, see Mikalson 1972.

**Figure 11:** The revised eclipse events' distribution on the cells of the Saros spiral (according **Table 3**). The Saros cells were re-numbered according to Voulgaris et al., 2021. The solar eclipse events classification is distributed on the bottom right half of the Back plate and the lunar eclipse events on the bottom left half according to the preserved BPI, and following the eclipse magnitude words' pattern μ]ικϱαί-μέσαι-με[γά]λ(αι)-μέσαι-μικϱαί. The phrase Η ΤΟΥ ΗΛΙΟΥ ΜΕΓΙΣΤΗ ΕΓΛΕΙΨΙΣ (or simply the word ΜΕΓΙΣΤΗ) + index letter A and afterwards the word ΜΕΙΖΟΝΕΣ are engraved as headings on the right BPI column, and Η ΤΗΣ ΣΕΛΗΝΗΣ ΤΕΛΕΙΑ (or ΜΕΓΙΣΤΗ) ΕΓΛΕΙΨΙΣ (+ index letter Γ2 - Γ circumflex) and afterwards the word ΜΕΙΖΟΝΕΣ are engraved as the headings of the left BPI column. The eclipse events sequence starts with *The Greatest Solar Eclipse* (New Moon at Node and at Apogee, see Voulgaris et al., 2023a, 2023b and 2023c) on the last day of month/Cell-1. After one Sar (half Saros) the Lunar eclipse event on Cell-113 occurs exactly at Node-A and therefore it is a Total Lunar eclipse, perfectly centered to the Earth's shadow (Full Moon at Perigee and at Node, Voulgaris et al., 2021 and 2023a; Meeus 1997, p.110-112), named *The Perfect Lunar Eclipse*. The 62th event (solar eclipse) 2 [A3] is on Cell-212 (49th cell with event), and the 63/64th events on Cell-218 ω[B3] (50th cell with event) (see **Table 3**).

The Antikythera Mechanism was a time measuring device based on the timed lunar cycle(s). Via the motion of the gears and pointers located inside a wooden box, this ΧΡΟΝΟΥ ΑΥΤΟΜΑΤΟΝ (time Automaton, see Voulgaris et al., 2025a, p.270), was provided timed information for the Moon and the Sun on the sky/Ecliptic zone, since both celestial objects regulated the social and the religious life in ancient Greece, as well as the most important astronomical events.

## APPENDIX-A

Based on the more precise equation by Chaldean astronomers the results show that One Saros ≈ 241.998 Draconic cycles. The impact due to this value (instead of 242), which is not an integer number (= 1.996 π), is presented in Figures A1 and A2.

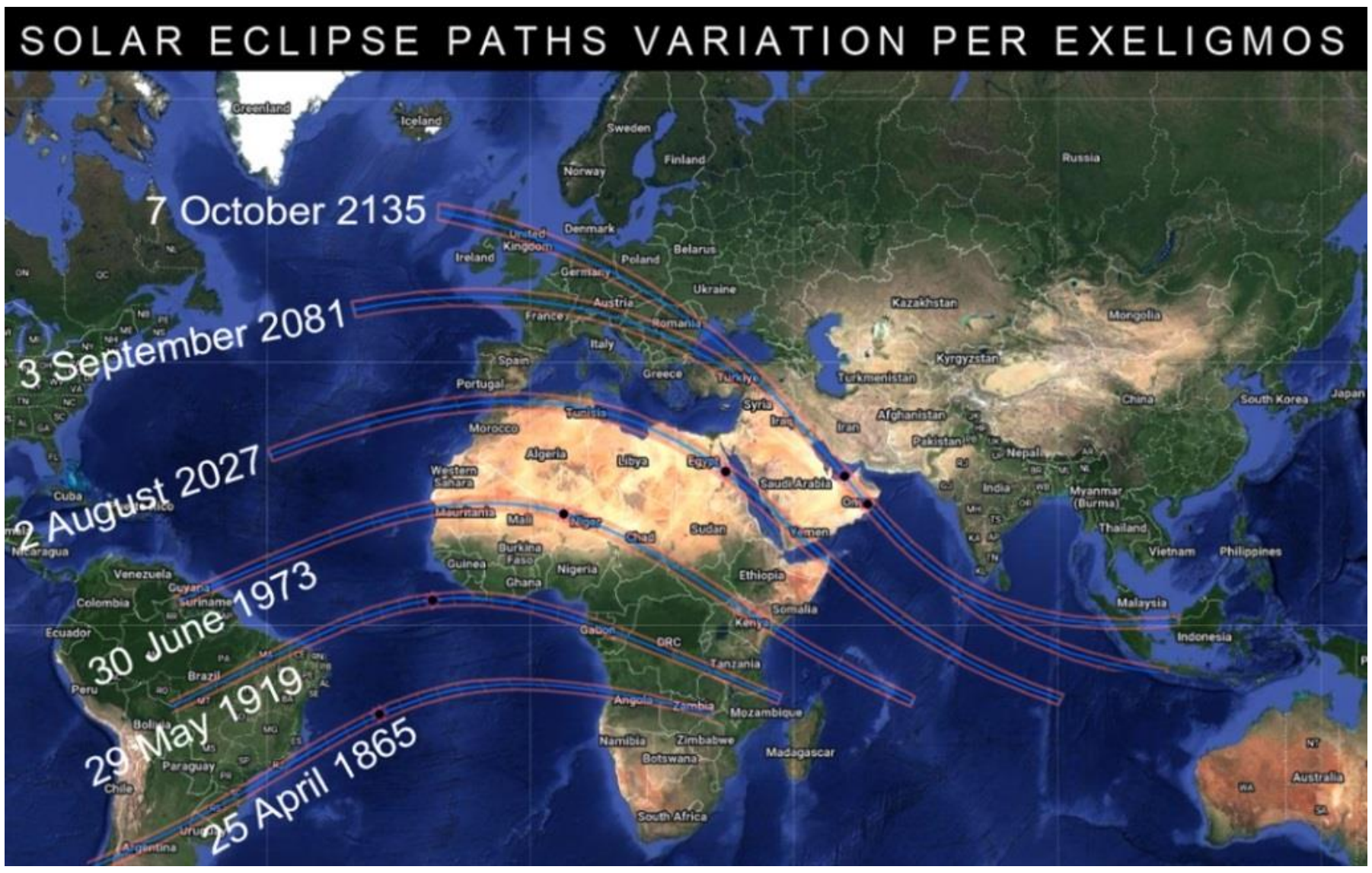

**Figure A1:** The eclipse path gradually changes its position in every successive Exeligmos: the Moon locates northward per each Exeligmos cycle. If the number of Draconic cycles per/Exeligmos was an integer number, the Moon's position should be the same per each Exeligmos. Graphics by X.Jubier (http://xjubier.free.fr/en/site_pages/Solar_Eclipses.html).

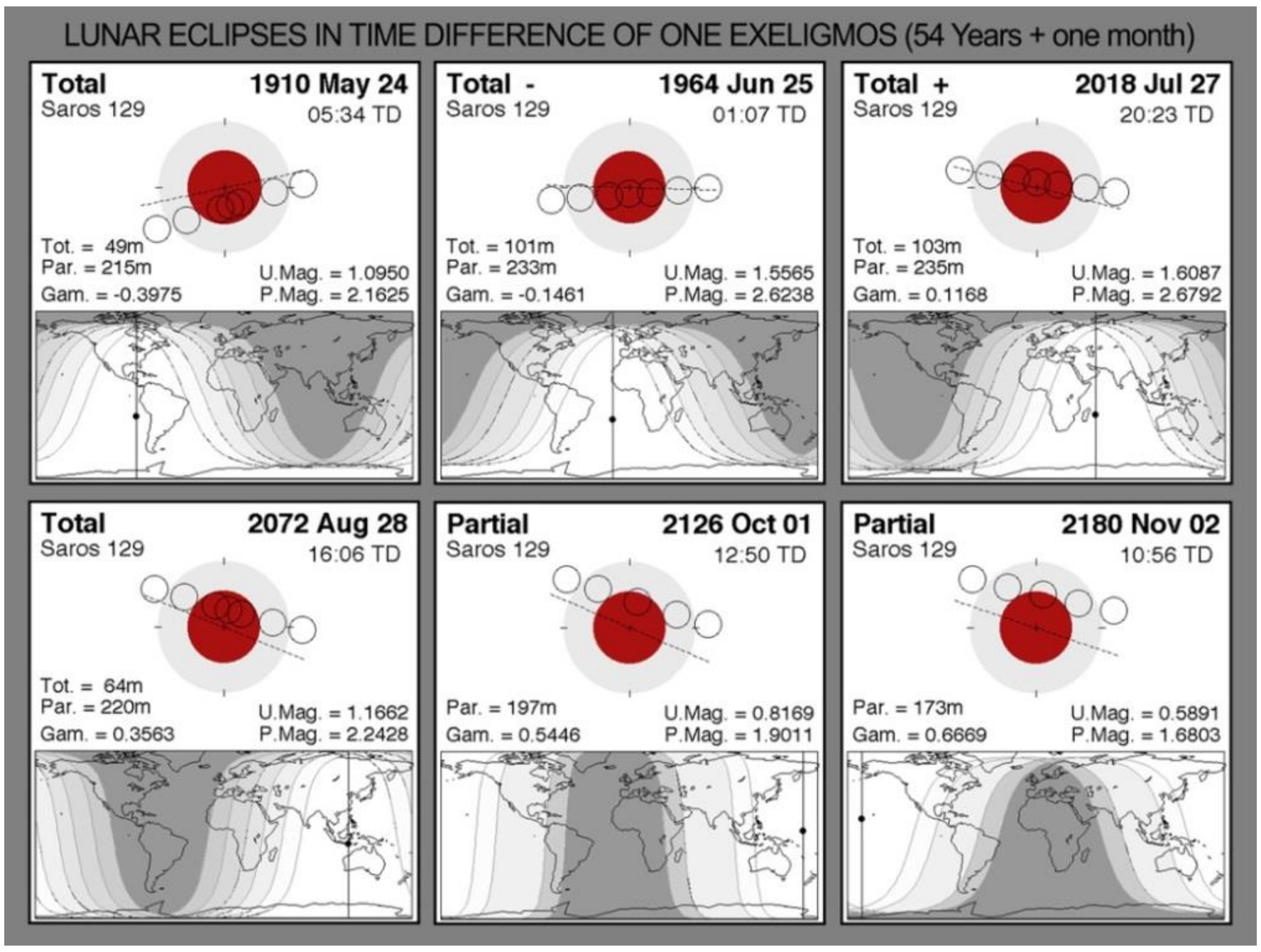

**Figure A2:** The lunar path gradually changes its position in every successive Exeligmos (also Saros). During time, the Moon is directed to the Northern (or Southern) part of the Earth's umbra, and the type of the lunar eclipse changes. This is the impact of the equation 669 synodic cycles ≈ 725.994 draconic cycles (instead of integer 726). Graphics by NASA Eclipse Web Site, "Five Millennium catalog of Lunar eclipses", by F. Espenak, NASA GSFC, https://eclipse.gsfc.nasa.gov/LEcat5/LEcatalog.html.

Using the revised *DracoNod* program (Voulgaris et al., 2023b) we present the eclipse events which correspond to the preserved classification index letters on the BPI of fragments A and F (Freeth 2014 and 2019; Anastasiou et al., 2016b; Iversen and Jones 2019; Jones 2020).
The ancient astronomers directly correlated the Draconic cycle phase to the eclipse magnitude (today, the Anomalistic phase is also included, as it also affects the eclipse magnitude). The program starts with the end of Cell-1, and the three lunar cycles (Synodic, Draconic and Anomalistic) are set on the New Moon at Node-A (and at Apogee). Then, according to **Table 3**, the preserved classified solar eclipse events are detected. *DracoNod* presents the Draconic cycle phase of the New Moon (and Full Moon) for the corresponding index letters of the events.

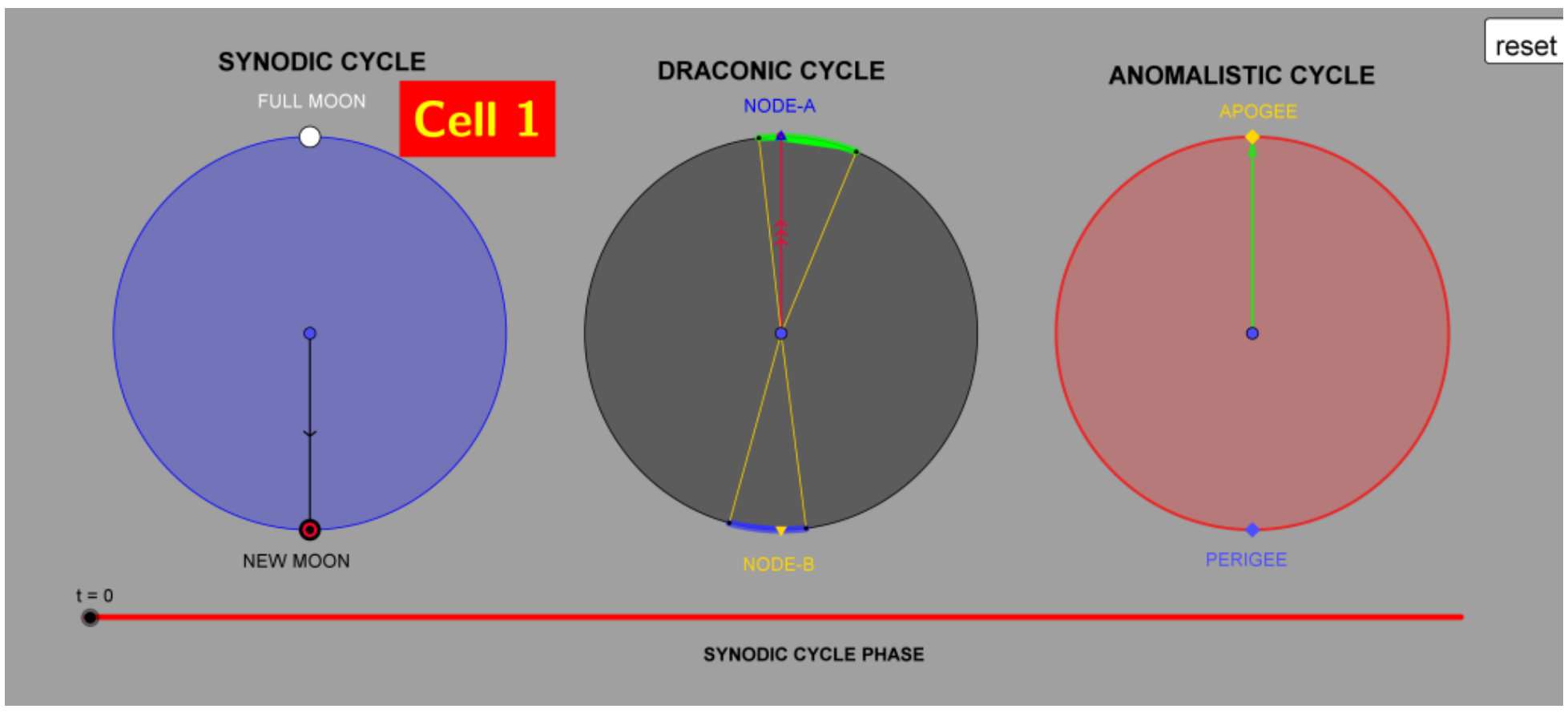

**Figure A3:** DracoNod starts with New Moon at Node-A and at Apogee.

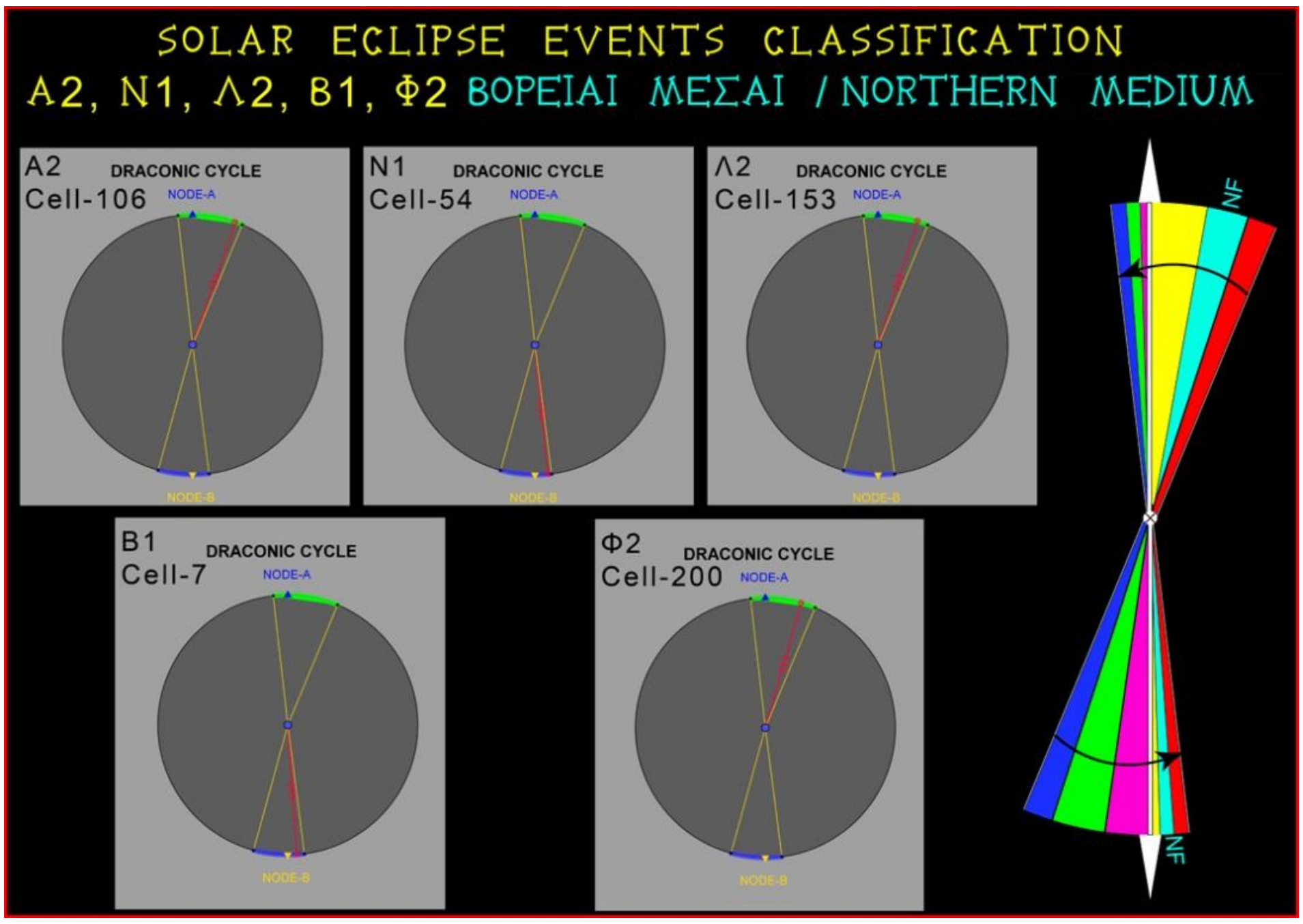

**Figure A4: ΒΟΡΕΙΑΙ ΜΕΣΑΙ ΕΓΛΕΙΨΕΙΣ** (**Northern Medium eclipses**, the Moon is North and Far from the Area of the Node, **Sector-NF**). The position of the New Moon relative to a Node (Draconic cycle phase red radius), for the solar eclipse events' preserved classification A2, N1, Λ2, B1, Φ2 (BPI L.1-9 in Iversen and Jones 2019) calculated by *DracoNod-V2*. These five events occurred very close to the Northern limit(s) of the two Nodes. The minor positioning mismatches can be justified by the gearing errors.

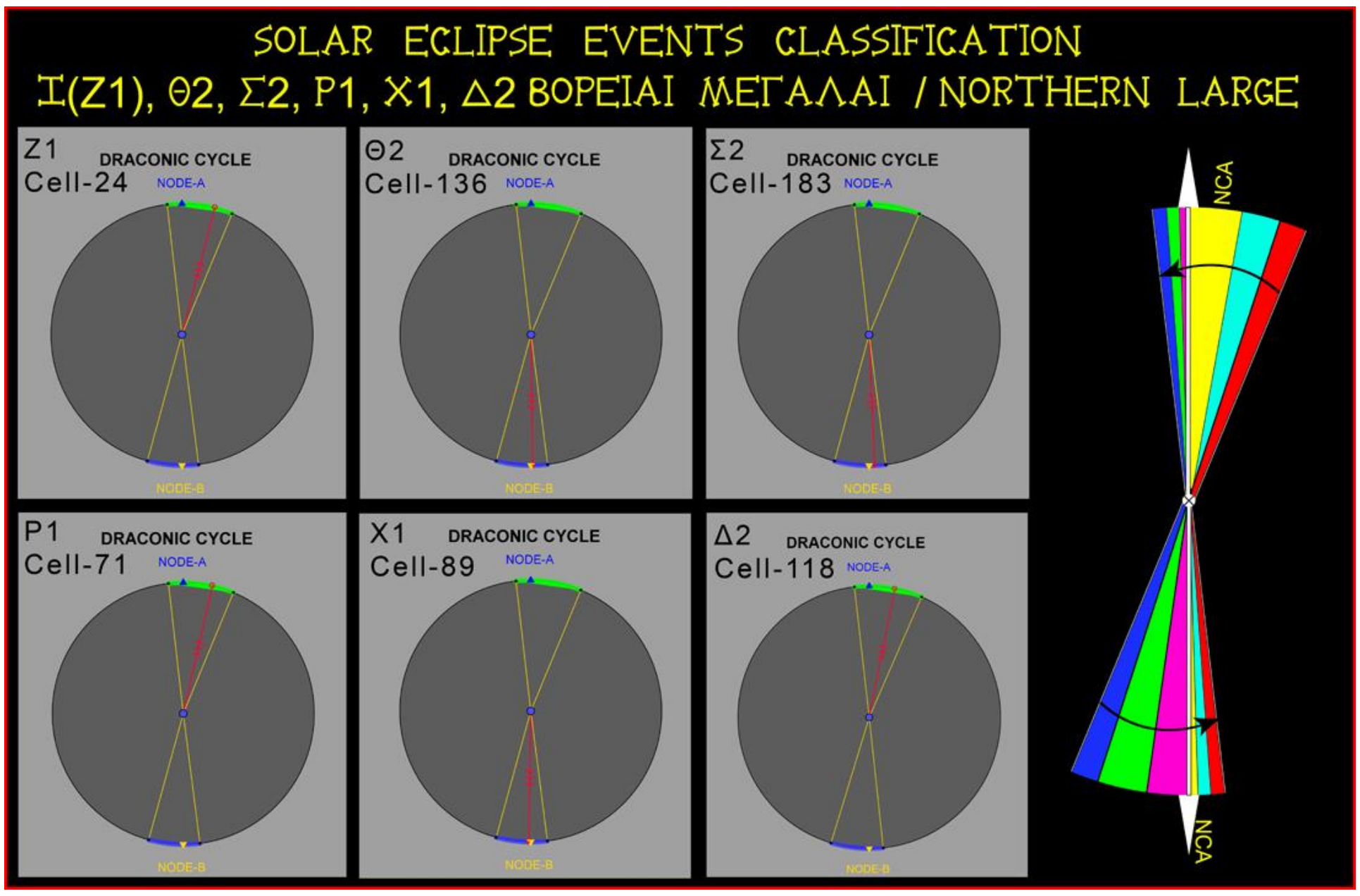

**Figure A5: ΜΕΓΑΛ<ΑΙ> ΕΓΛΕΙΨΕΙΣ** (**Large eclipses**, the Moon is just out of the Area of the Node, **Sector-NCA**). The position of the New Moon relative to a Node (Draconic cycle phase red radius) for the solar eclipse events' preserved classification Z1, Θ2, Σ2, P1, X1, Δ2 (BPI L.10-20 Iversen and Jones 2019), calculated by DracoNod-V2. The events Θ2, Σ2, X1 occurred close to the Node. The rest events Z1, P1, Δ2 occurred North and far from the Node. These mismatches on the positioning can be justified by the gearing errors.

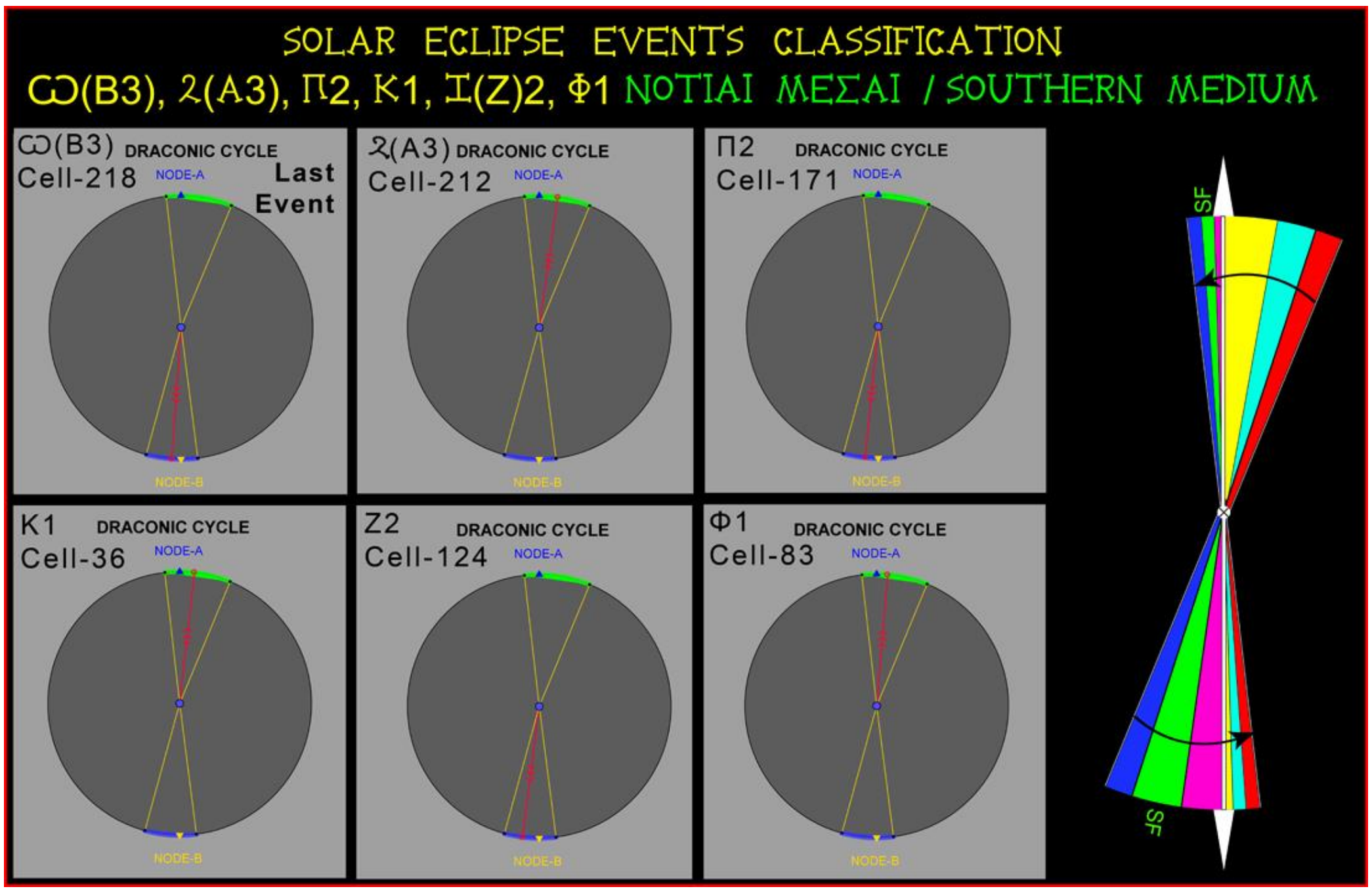

**Figure A6:** BPI L.21-31(Iversen and Jones 2019), **NOTIAI ΜΕΣΑΙ ΕΓΛΕΙΨΕΙΣ** (**Southern Medium eclipses**, the Moon is south and far from the Area of the Node, **Sector-SF**). The position of the New Moon relative to a Node (Draconic cycle phase red radius), for the solar eclipse events' preserved classification Ꞷ/B3, Ꝛ/A3, Π2, K1, Z2, Φ1 calculated by *DracoNod-V2*. The events Ꝛ/A3, K1, Φ1, occurred North and out of the Node-A. The events Ꞷ/B3, Π2, Z2 occurred South and out of the Node. The mismatches on the pointer's positioning can be justified by the gearing errors.

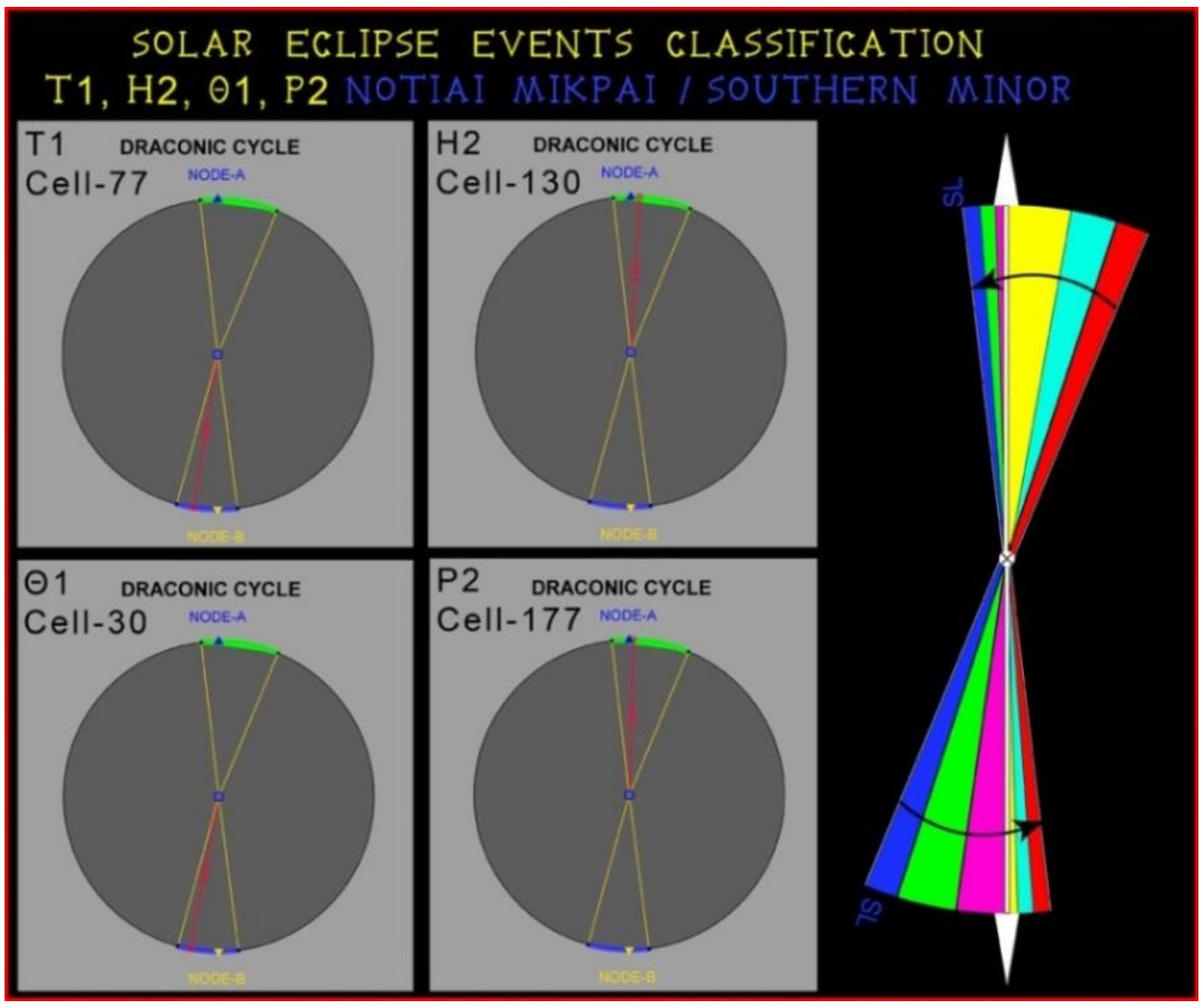

**Figure A7:** BPI L.32-38 (Iversen and Jones 2019), **NOTIAI ΜΙΚΡΑΙ ΕΓΛΕΙΨΕΙΣ** (**Southern Minor eclipses**, the Moon is Close to/On Southern Limit, **Sector-SL**).The position of the New Moon relative to a Node (Draconic cycle phase red radius), for the solar eclipse events' preserved classification T1, H2, Θ1, P2 calculated by DracoNod-V2. The events T1, Θ1, occurred close to the southern ecliptic limit. The mismatches on the pointer's position can be justified by the gearing errors.

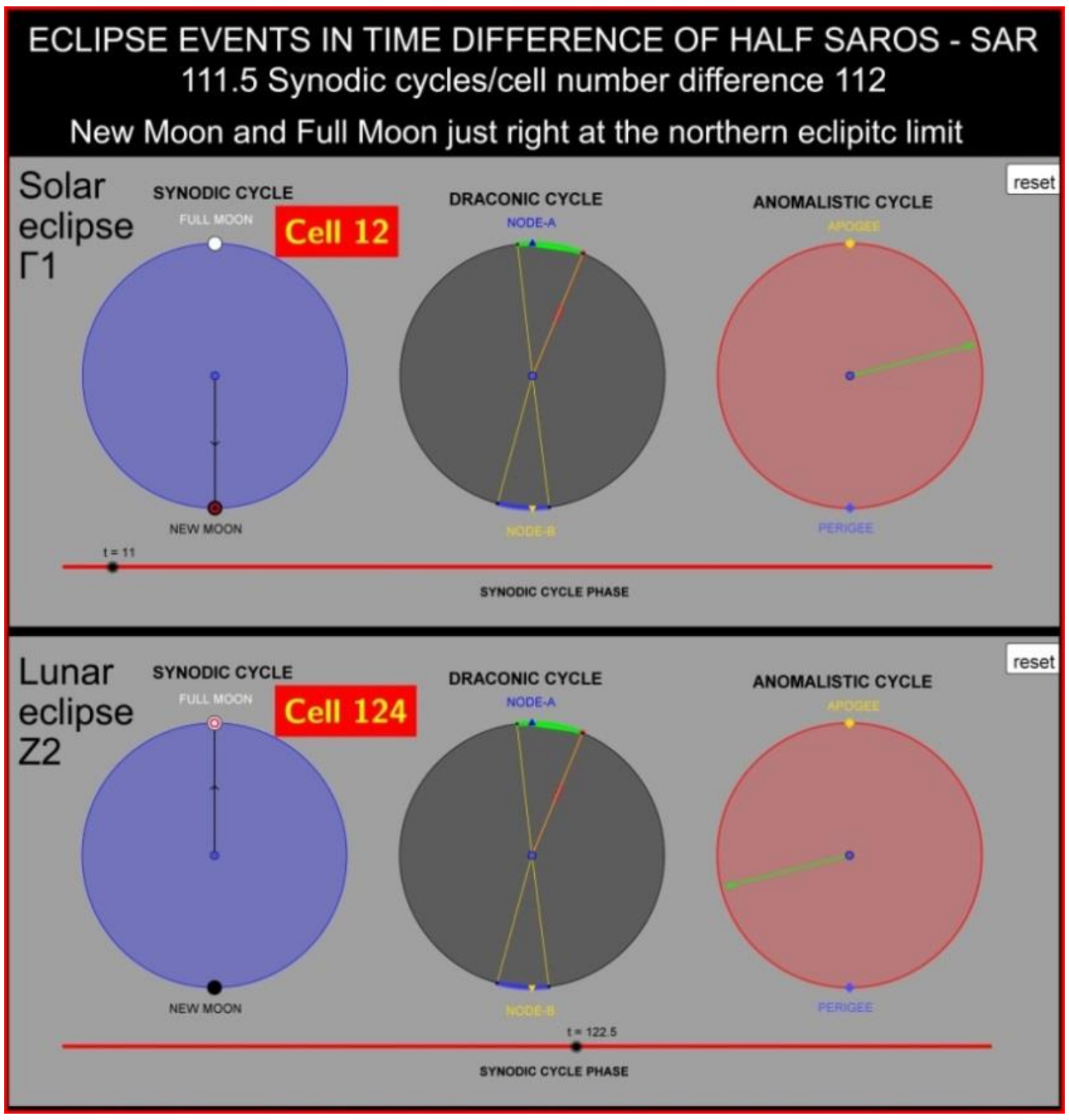

**Figure A8:** Two eclipse events separated by one Sar (half Saros) period: a solar eclipse (event Γ1 on Cell-12) and a lunar eclipse (event Z2 on Cell 124). Both events occurred exactly on the northern ecliptic limit of Node-A (see the red pointer of Draconic scale). These two events lead to the conclusion that the ancient Craftsman used common ecliptic limits for the solar and lunar eclipses.

**Table A:** After one Sar period, the eclipse events repeat in inverted order (solar/lunar eclipse), at (about) the same ecliptic latitude and in inverted anomalistic phase (apogee/perigee), see **Figure A8**.

| **INVERTED ECLIPSE EVENTS IN TIME DIFFERING BY A HALF SAROS/SAR<br>111.5 Synodic cycles/121.0 Draconic cycles/119.5 Anomalistic cycles** | | |
|---|---|---|
| **New Moon** | **After one Sar period ≈ $9^{Y}\,5.5^{d}$** | **Full Moon** |
| At Node-A | | At Node-A |
| At Apogee | | At Perigee |
| x° North/South of Node-A/B | | x° North/South of Node-A/B |
| At Northern/Southern ecliptic limit | | At Northern/Southern ecliptic limit |
| At Node A/B + at Apogee = Annular solar eclipse (longest duration) | | At Node A/B + at Perigee = Total lunar eclipse (shortest duration) |
| At Node-A/B + at Perigee = Total solar eclipse (longest duration) | | At Node-A/B + at Apogee = Total lunar eclipse (longest duration) |
| Several x-degrees North/South of Node-A/B = Total/Annular solar eclipse | | Several x-degrees North/South of Node-A/B = Partial lunar eclipse |
| At North/South ecliptic limit = partial solar eclipse visible from North/South pole | | At North/South ecliptic limit = penumbral lunar eclipse North/South of Earth's shadow |

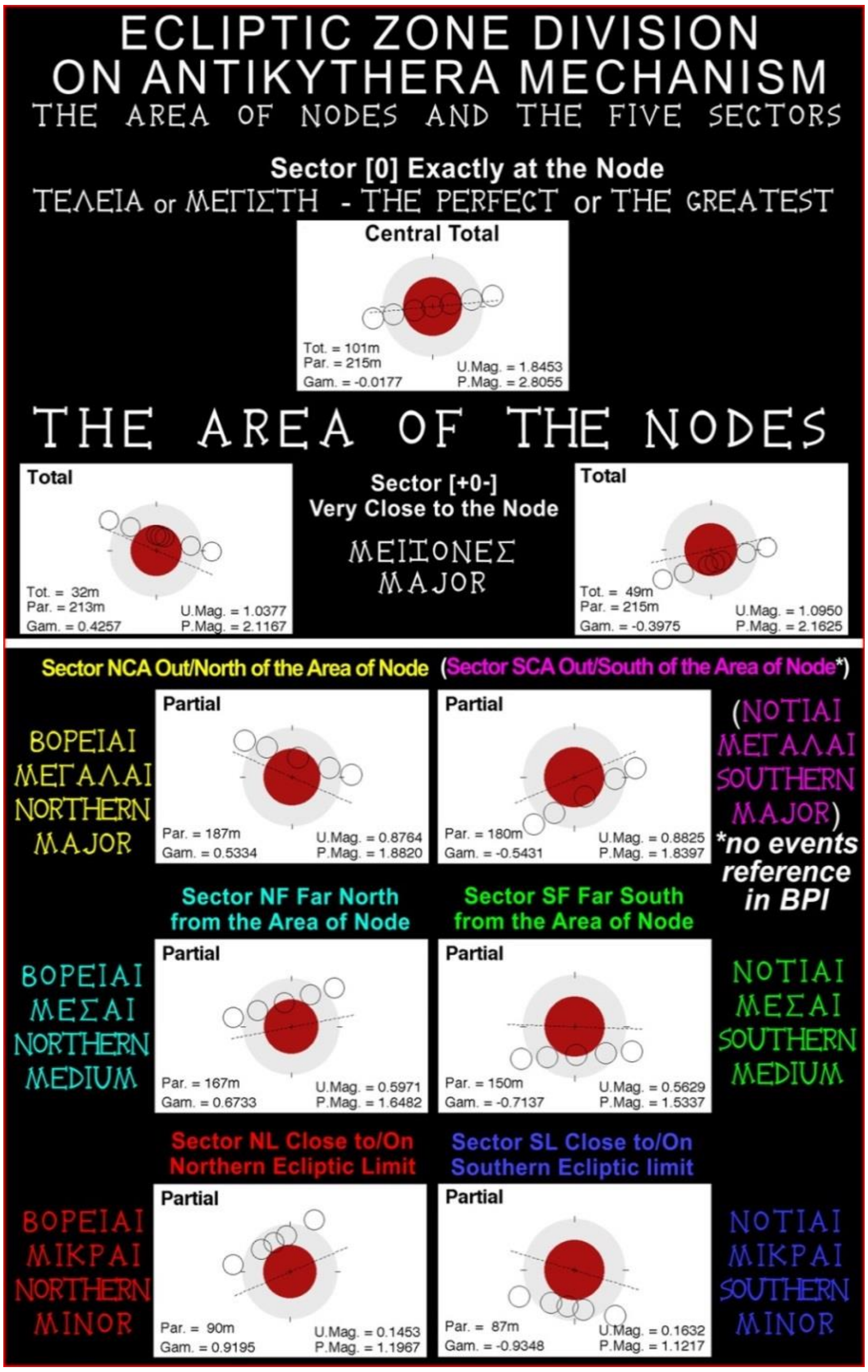

**Figure A9:** Division of the Ecliptic Zone on the Area of Node and into six Sectors. The specific division is based on the retracted data from Eudoxus papyrus, the BPI of the Antikythera Mechanism and on the observational characteristics. The eclipses happening on the Area of Node (Perfect or Major) are central total or about central total or total at the limit. During a central total eclipse, the lunar disc displays a homogenous brightness. During the non-central total lunar eclipses (i.e. a bit out of Node, but inside the Earth's umbra), a bright arc at the contact boundary with the umbra is visible. This is the characteristic difference between a total central eclipse (Τελεία or Μεγίστη-The Greatest) and a non-central total one (Μείζονες-Major). This difference is easily detectable by an observer. The arrangement of the images follows the pattern of the Antikythera Mechanism BPI and the lost data retracted by Eudoxus papyrus. Selected graphics by the *Index to five millennium Catalog of Lunar eclipses by NASA* (https://eclipse.gsfc.nasa.gov/LEcat5/LEcatalog.html).

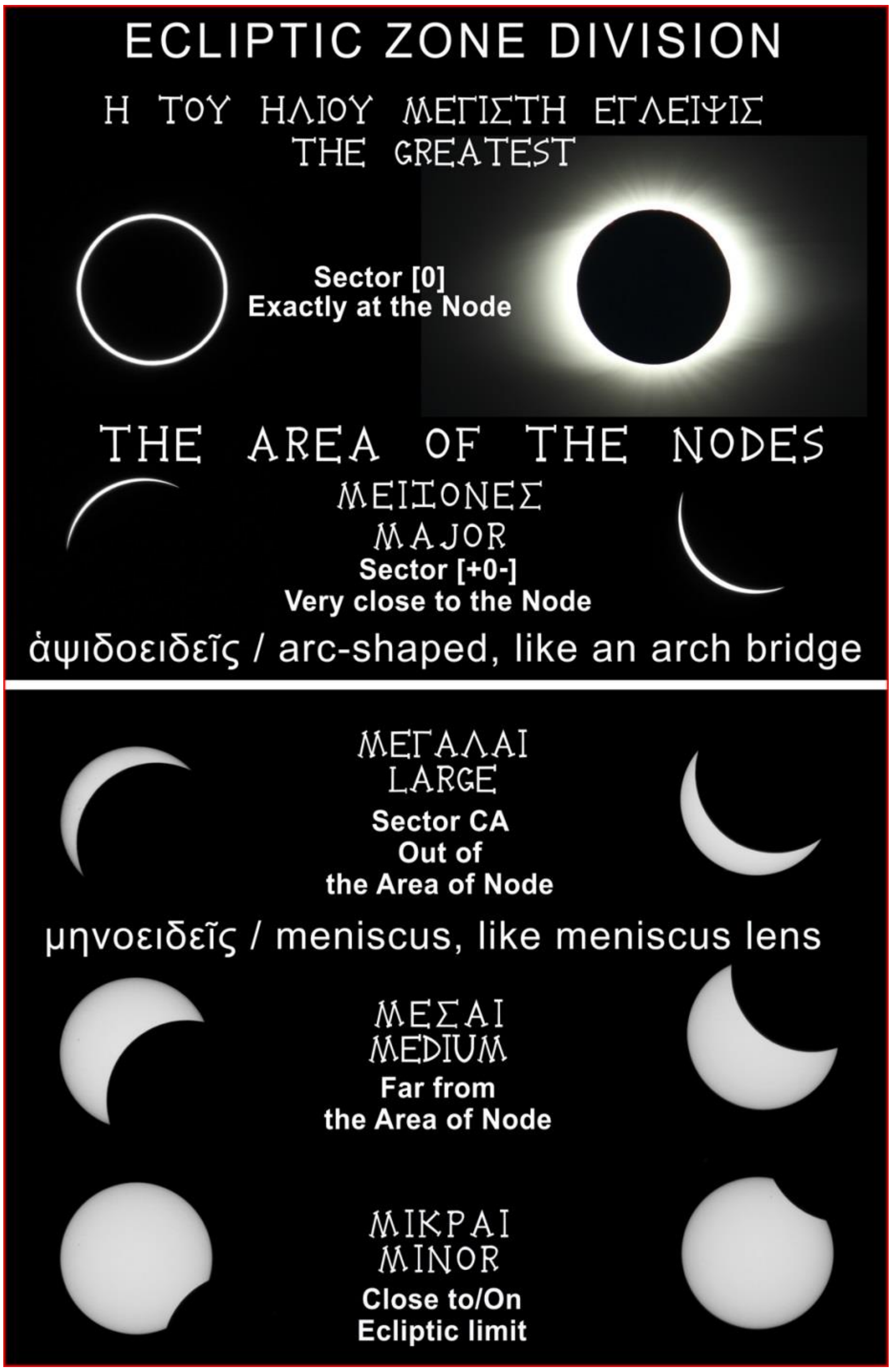

**Figure A10:** The eclipse classification based on the Eclipse magnitude/Obscuration of the solar disc by the lunar disc. On top, the ΜΕΓΙΣΤΗ ΗΛΙΟΥ ΕΓΛΕΙΨΙΣ/The Greatest eclipse, which is either total or annular. In Eudoxus papyrus (col. XX, 9-13) is mentioned that οὐχ ὅλοωι τῶι ἡλίωι ἐπισκοτεῖ ἐν τῆι μεγίστηι τοῦ ἡλίου ἐγλείψει. Μείζων ἄρα ἔσθ’ ὁ ἥλιος τῆς Σελήνης, during the Greatest solar eclipse, the Moon does not cover the full disc of the Sun, therefore the Sun is larger than the Moon): this is a description of an annular solar eclipse occurring exactly at the Node. Cleomedes (II, 4 in Ziegler 1891) writes about the Τελεία Ἡλίου ἔκλειψις (Perfect solar eclipse): … ἐν ταῖς τελείαις τῶν ἐκλείψεων, ὅτε ἐπὶ μιᾶς εὐθείας γίνεται τὰ **κέντρα τῶν θεῶν**, κύκλῳ περιφαίνεσθαι πάντοθεν ἐξέχουσαν τὴν ἴτυν τοῦ ἡλίου, the Perfect solar eclipses occur when the **Centers of Gods**-the centers of the three celestial bodies) are aligned in a straight line, and a circle of the Sun (ring) remains uncovered by the Moon, (Cleomedes describes an annular solar eclipse). He also states: Αἱ μέν ἐλάσσους μηνοειδεῖς (the lesser eclipses appear as meniscus), αἱ δέ μείζους ἀψιδοειδεῖς (the major eclipses appear in arc-shape/like a bridge), αἱ δέ μείζους (Σελήνης ἐγλείψεις) ᾠοειδεῖς (the major lunar eclipses appear in oval shape). The ΜΕΣΑΙ/Medium eclipses correspond to about 50% coverage. The ΜΙΚΡΑΙ/Minor eclipses are those that only a small part of the solar disc is covered by the lunar disc. The photos were taken by the first author during: the annular solar eclipse on 26th December 2019 from Sharqiya Desert, Oman, the total solar eclipse of 2nd July 2019 from Cerro Tololo Inter-American Observatory in Atacama Desert, Chile, and the partial solar eclipse phases of 29th March 2006 from Kastellorizo Island, Greece.

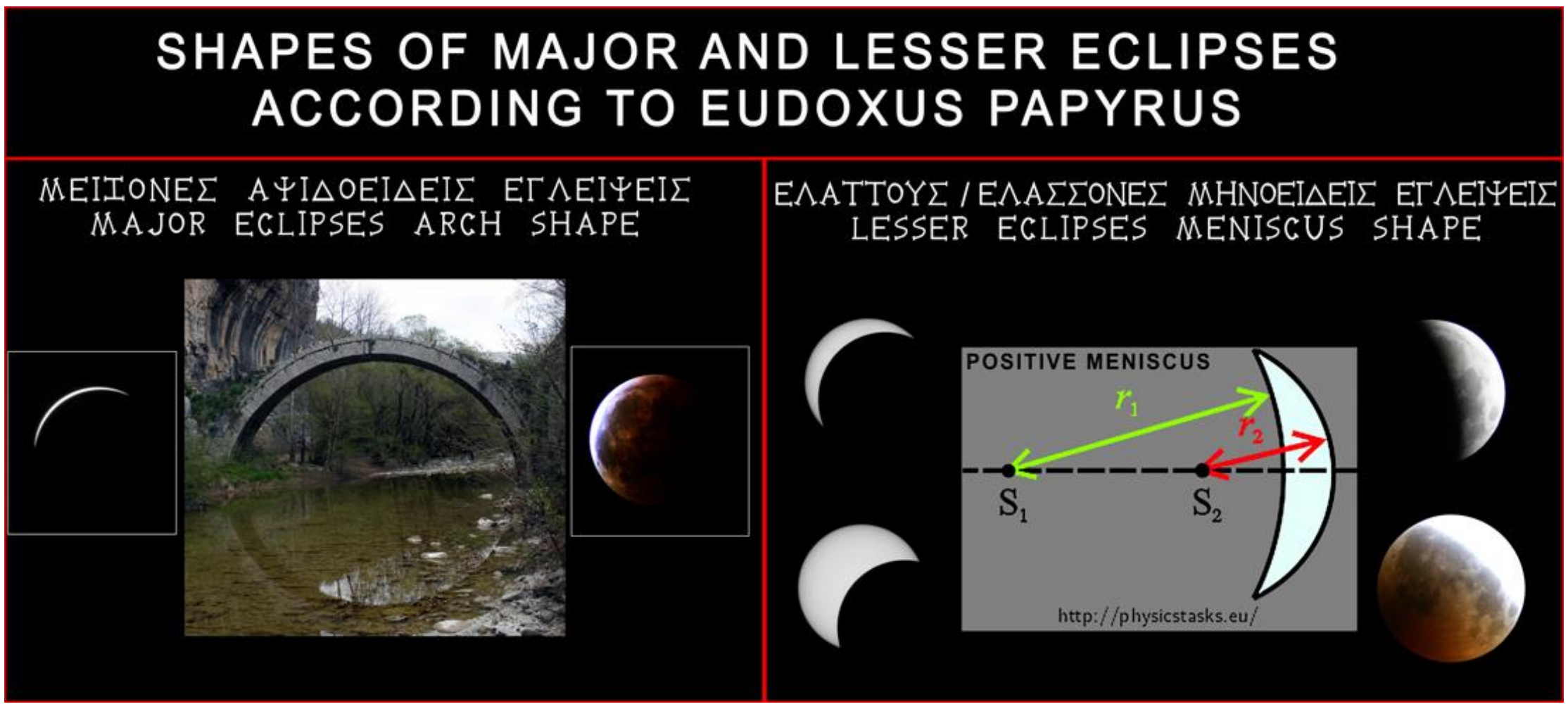

**Figure A11:** Left, the ΑΨΙΔΟΕΙΔΕΙΣ eclipses are arch-shape. Insert: The stone Bridge of Kontodimos or Lazaridis, in arch shape design (located between the villages of Kipi and Koukouli, in Epirus, NW Greece, photo by A.Vossinakis). In this design, the inner and the outer arcs of the bridge seem parallel to each other. Right, the shape of the ΜΗΝΟΕΙΔΕΙΣ eclipses seem like a positive meniscus lens. Insert: positive meniscus optical scheme. Note the difference between the radii of curvature of the two optical surfaces, credits https://physicstasks.eu/2216/positive-meniscus-lens). The photographs of the eclipsed Sun and the eclipsed Moon were captured by first author on different dates.

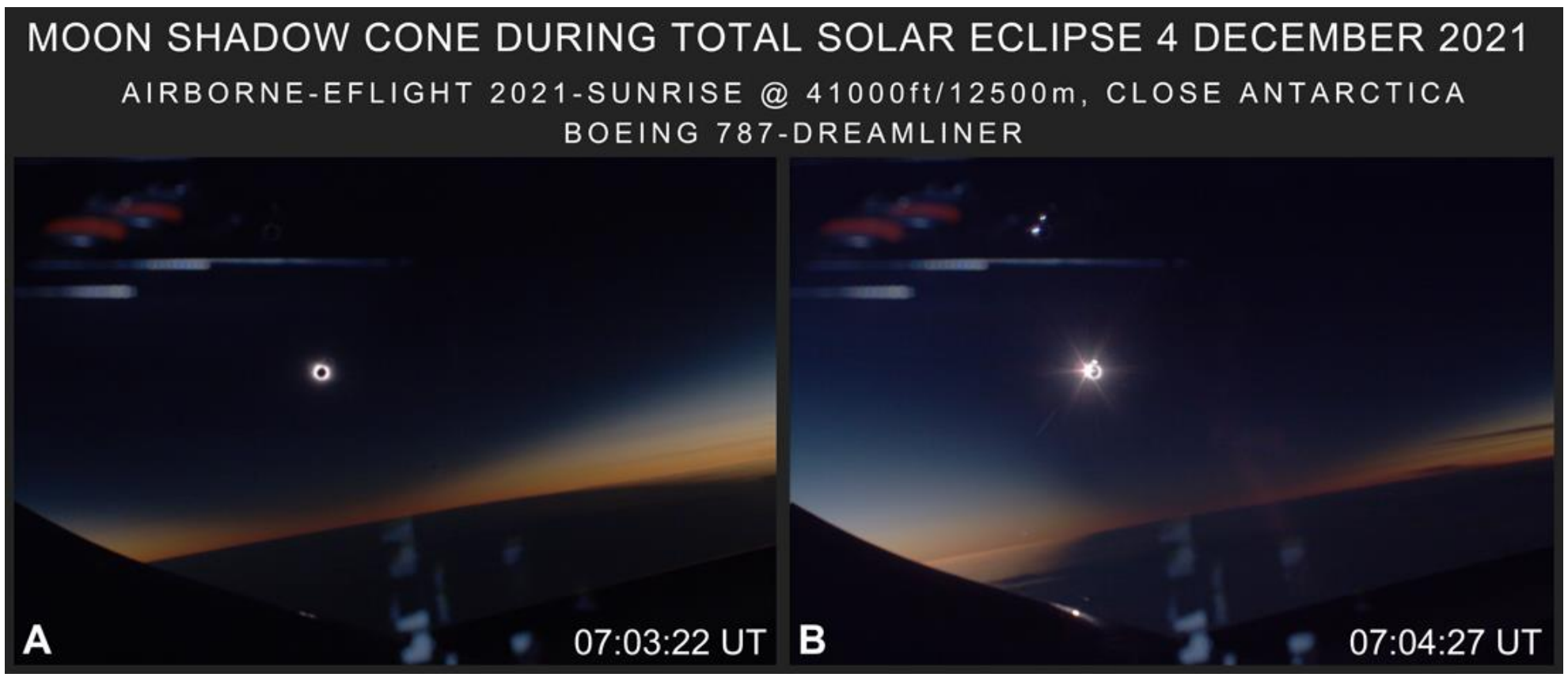

**Figure A12:** A modern recording of the Moon shadow during the total solar eclipse on 4$^{th}$ December 2021, close to Antarctica. The photographs were taken by the first author during the airborne solar eclipse expedition EFLIGHT 2021-SUNRISE from an altitude of 41000ft/12500m (http://nicmosis.as.arizona.edu:8000/ECLIPSE_WEB/TSE2021/TSE2021WEB/EFLIGHT2021.html). A) Close to the eclipse maximum. B) Taken at 3$^{rd}$ contact/end of totality. The solar eclipse was visible from the very southern parts of the Earth, since the New Moon was very close to the southern ecliptic limit (see also **Figure 9C**). The shadow cone transits the Earth's surface with a hypersonic velocity of ≈ 6.3 km/sec (≈22,680 km/h ≈ 19 Mach).

## Acknowledgements

We are very grateful to Professors M. Edmunds (Cardiff University, UK), J. Seiradakis (Aristotle University, Thessaloniki, GR) and X. Moussas (National and Kapodistrian University of Athens, GR), who provided us with the X-Ray raw volume data of the Antikythera Mechanism's fragments and Dr. F. Ullach for his support in the use of the REAL3D VOLVICON Software. Thanks are due to the National Archaeological Museum of Athens, Greece, for permitting us to photograph the Antikythera Mechanism fragments. We would like to thank Prof. Zach Ioannou of Sultan Qaboos University of Muscat, Oman, for the hospitality and support to observe/record via first author's coronagraphs and spectrographs the Annular Solar Eclipse of December 26$^{th}$ 2019, from Sarqiya Desert and also Prof. Tom Economou of Fermi Institute-University of Chicago, USA, for his help on the preparation of the eclipse expedition and observation. The first author has participated in 13 solar eclipse research expeditions under Prof. Dr. J.M. Pasachoff's research, performing spectroscopic observations of the solar corona. For Total Solar Eclipses of 2017 (USA), 2019 (Chile, Cerro Tololo Inter-American Observatory-Atacama Desert), 2021 (airborne solar eclipse observation EFLIGHT–SUNRISE 2021) and 2023 (Australia) under Prof. Dr. J.M. Pasachoff's research, sponsored by Grant AGS–1903500 of the Solar Terrestrial Program, Atmospheric and Geospace Sciences Division of the U.S. National Science Foundation, succeeding AGS–1602461.
The authors' first paper describing the idea for the Draconic gearing existence on the Antikythera Mechanism was initially submitted to a journal on January 14, 2020, and it was eventually published in a different journal (*Mediterranean Archaeology and Archaeometry*) on December 2022.